\documentclass[twocolumn,twocolappendix,times]{aastex6}
\usepackage{amsmath}
\usepackage{amssymb}
\usepackage{xspace}
\usepackage{xfrac}
\usepackage{xcolor}
\definecolor{green}{HTML}{228B22}

\newcommand{\Reff}{R_{\mathrm{eff}}\xspace}

\newcommand{\Rhalo}{R_{200c}\xspace}
\newcommand{\Mstellar}{M_*\xspace}
\newcommand{\Mhalo}{M_{200c}\xspace}

\newcommand{\HST}{\emph{HST}\xspace}

\begin{document}

\title{Relations Between the Sizes of Galaxies and their Dark Matter Halos at Redshifts $0 < \lowercase{z} < 3$}

\author{Kuang-Han Huang\altaffilmark{1}}
\author{S. Michael Fall\altaffilmark{2}}
\author{Henry C. Ferguson\altaffilmark{2}}
\author{Arjen van der Wel\altaffilmark{3}}
\author{Norman Grogin\altaffilmark{2}}
\author{Anton Koekemoer\altaffilmark{2}}
\author{Seong-Kook Lee\altaffilmark{4}}
\author{Pablo G. P\'erez-Gonz\'alez\altaffilmark{5}}
\author{Stijn Wuyts\altaffilmark{6}}

\altaffiltext{1}{University of California Davis, 1 Shields Avenue, Davis, CA 95616, USA; khhuang@ucdavis.edu}
\altaffiltext{2}{Space Telescope Science Institute, 3700 San Martin Drive, Baltimore, MD 21218, USA}
\altaffiltext{3}{Max Planck Institute for Astronomy, Koenigstuhl 17, D-69117 Heidelberg, Germany}
\altaffiltext{4}{Center for the Exploration of the Origin of the Universe, Department of Physics and Astronomy, Seoul National University, Seoul, Korea}
\altaffiltext{5}{Departamento de Astrof\'isica, Facultad de CC. F\'isica, Universidad Complutense de Madrid, E-28040, Madrid, Spain}
\altaffiltext{6}{Department of Physics, University of Bath, Claverton Down, Bath, BA2 7AY, UK}

\email[E-mail:~]{khhuang@ucdavis.edu}

\begin{abstract}
We derive relations between the effective radii $R_{\rm{eff}}$ of galaxies and the
virial radii $R_{200c}$ of their dark matter halos over the redshift range $0 < z < 3$. 
For galaxies, we use the measured sizes from deep images taken with \emph{Hubble Space Telescope}
for the Cosmic Assembly Near-infrared Deep Extragalactic Legacy Survey; for halos,
we use the inferred sizes from abundance matching
to cosmological dark matter simulations via a stellar mass--halo mass (SMHM) relation. 
For this purpose, we derive a new SMHM relation based on the same selection criteria 
and other assumptions as for our sample of galaxies with size measurements. As a check
on the robustness of our results, we also derive $R_{\rm{eff}}$--$R_{200c}$ relations
for three independent SMHM relations from the literature.
We find that galaxy $R_{\rm{eff}}$
is proportional on average to halo $R_{200c}$, confirming and extending to 
high redshifts the $z=0$ results
of Kravtsov. Late-type galaxies (with low S\'ersic index and high specific
star formation rate [sSFR]) follow a linear $R_{\rm{eff}}$--$R_{200c}$ relation, with
effective radii at $0.5 < z < 3$ close to those predicted by simple models
of disk formation; at $z < 0.5$, the sizes of late-type galaxies
appear to be slightly below this prediction.
Early-type galaxies (with high S\'ersic index and low sSFR) follow a 
roughly parallel $R_{\rm{eff}}$--$R_{200c}$ 
relation, $\sim$ 0.2--0.3 dex below the one for late-type galaxies.
Our observational results, 
reinforced by recent hydrodynamical simulations, indicate that galaxies grow
quasi-homologously with their dark matter halos.

\end{abstract} 


\section{Introduction}

The size of a galaxy, as measured by its half-mass radius $R$, for example,
is among the most basic of its properties.  Together with the mass $M$,
the size $R$ determines the binding energy, $-E\approx{GM^2}/(4R)$, and hence 
the energy radiated away during the formation of the galaxy.  For galactic disks, 
with stars and gas on nearly circular orbits with rotation velocity $V_{\rm{rot}}$, 
the size $R$ is determined by the angular momentum $J\approx{MRV_{\rm{rot}}}$, 
which in turn determines the energy $E=-\mathrm{\sfrac{1}{2}}MV_{\rm{rot}}^2\approx-{G^2M^5}/(8J^2)$. 
The basic description of galaxies in general consists of $M$, $R$, and $V_{\rm{rot}}$, 
or equivalently $M$, $E$, and $J$, while for disk-dominated galaxies, any two of 
these quantities suffice.      

As a result of the hierarchical growth of galaxies, we expect their masses and 
radii to increase with cosmic time and thus to decrease with redshift.  In the 
simplest models of galaxy formation, the sizes of the baryonic components of galaxies are, 
on average, proportional to the sizes of their surrounding dark matter halos. 
For galactic disks, this proportionality in sizes follows directly from the assumed 
proportionality of the specific angular momentum of baryons and dark matter resulting from 
tidal torques in the early stages of galaxy formation \citep{Fall:1980up,Mo:1998hg}. 
This assumption underlies practically all of the semianalytical models of galaxy formation 
in current use \citep[e.g.,][]{Cole:2000fl,Croton:2016jg}. Recent hydrodynamical simulations 
of galaxy formation confirm the approximate proportionality between the specific angular momentum 
of galaxies and their dark matter halos \citep{Genel:2015kp,Pedrosa:2015dh,Teklu:2015ev,Zavala:2016ki}.

There have been numerous searches for the expected decrease in galactic sizes with redshift 
based on measurements of deep images taken with the \emph{Hubble Space Telescope} (\HST) over the 
past dozen years \citep[e.g.,][]{Ferguson:2004dt,Hathi:2008ca,Mosleh:2012cw}. These searches 
all find that galaxies were smaller in the past, by roughly the predicted amount, 
although there are significant differences in the precise decline of galactic sizes with 
redshift among these studies (compare, e.g. \citealt{Shibuya:2015bj} and \citealt{CurtisLake:2016bm}). 
Part of the discrepancy among these results stems from the fact that the apparent evolution in 
sizes depends on how galaxies at different redshifts are compared, whether at fixed 
stellar mass or luminosity or at variable stellar mass or luminosity.

\cite{Kravtsov:2013cy} used stellar mass--halo mass (SMHM) relations derived via the 
technique of abundance matching to compare the observed sizes of present-day galaxies 
with the sizes of their matched dark matter halos in cosmological $N$-body simulations. 
He found that the sizes of galaxies at $z=0$ are proportional on average to the sizes of 
their halos. Furthermore, the coefficient of proportionality is consistent with a simple
model in which galactic disks grow with approximately the same specific angular momentum as their
halos until $z \sim 2$ and then stop growing after that. The question immediately arises whether 
the same or a different relation holds between the sizes of galaxies and their halos at 
high redshifts. The purpose of this paper is to answer this question.

The advantage of comparing the sizes of galaxies at multiple redshifts with 
the sizes of their matched halos at the same redshifts, as we do here, 
is that the results are then expressed directly in simple, physically meaningful terms.
This framework also helps to clarify the results of previous searches
for the evolution of galactic sizes.

There are already a couple of indications that the sizes of galaxies and their 
halos evolve in lockstep.  First, semiempirical models of galaxy formation that 
make this assumption agree better with deep \HST images than the same models with 
different assumptions about the evolution of galactic sizes \citep{TaghizadehPopp:2015hp}. 
Second, recent measurements of the sizes and rotation velocities of galactic disks 
at $1 < z < 3$ and $0.2 < z < 1.4$ indicate that they have approximately the same 
specific angular momenta 
as their dark matter halos \citep{Burkert:2016fr,Contini:2016fn}. While these results 
are suggestive, it is still important to make a direct, independent comparison of the 
sizes of high-redshift galaxies with the sizes of their matched halos, 
the investigation we describe here.

The plan for the remainder of this paper is the following.  In Section \ref{sec:data}, 
we describe our sample of galaxies and measurements of their sizes and other properties. 
In Section \ref{sec:matching}, we discuss the abundance-matching method and its implementation
with four different SMHM relations. In Section \ref{sec:results}, 
we present the results of our comparison of galaxy and halo sizes, and  
in Section \ref{sec:errors}, we discuss the uncertainties in these results.
We discuss some implications of our results in Section \ref{sec:disc}.
We show the connection between the galaxy size--halo size relation and the
more familiar galaxy size--stellar mass relation in an appendix.
All magnitudes quoted in this paper are in the AB system, and we assume the following 
cosmological parameters: $h=0.7$, $\Omega_m=0.27$, and $\Omega_\Lambda=0.73$.

\floattable
\begin{deluxetable}{lcccccc}
\tabletypesize{\small}
\tablecolumns{7}
\tablecaption{Galaxy Sample Sizes\label{tab:sample}}
\tablehead{\colhead{Redshift} & \colhead{Wide} & \colhead{Deep} & \colhead{HUDF} & \colhead{Total} & \colhead{${z_{\rm{med}}}$} & \colhead{$M_{*,\rm{low}}$\tablenotemark{a}} \\
 &  &  &  &  &  & \colhead{($M_\odot$)}}

\startdata
$0.0 < z < 0.5$  & 4388  &  923 &  50 &  5361 & 0.34 & $1.0\times10^7$ \\
$0.5 < z < 1.0$  & 9706  & 2435 & 116 & 12257 & 0.73 & $5.0\times10^7$ \\
$1.0 < z < 1.5$  & 6666  & 1395 & 113 &  8174 & 1.23 & $8.2\times10^7$ \\
$1.5 < z < 2.0$  & 5152  & 1224 &  90 &  6466 & 1.70 & $1.7\times10^8$ \\
$2.0 < z < 2.5$  & 2580  &  727 &  47 &  3354 & 2.23 & $2.1\times10^8$ \\
$2.5 < z < 3.0$  & 1483  &  497 &  54 &  2034 & 2.69 & $3.8\times10^8$ \\
\hline
All Redshifts  & 29975  &  7201 &  470 &  37646 & \nodata & \nodata \\
\enddata
\tablenotetext{a}{Typical stellar mass of the galaxies from HUDF with
$26.6$ mag $< H_{160} < 26.8$ mag
and near the median of each redshift bin. In the lowest redshift bin, we impose a hard cut
in stellar mass at $10^7\,M_\odot$.}

\end{deluxetable}

\section{Observations}\label{sec:data}


For this study, we need a galaxy sample with homogeneous data quality 
that enables accurate size measurements. \HST images are required because 
galaxies at $z>1$ are generally smaller 
than $1\arcsec$. We also need a galaxy sample 
with good constraints on redshifts, stellar masses, and star formation rates, so that 
we can connect galaxies to dark matter halos and distinguish star forming galaxies 
from quiescent galaxies. The Cosmic Assembly Near-infrared Deep Extragalactic Legacy 
Survey (CANDELS) is the best data set currently 
available for this study: all five CANDELS fields, covering $\approx 800$ arcmin$^2$ in total, 
have \HST images at optical and near-IR wavelengths with uniform quality 
\citep{Grogin:2011hx,Koekemoer:2011br}. The high 
angular resolution of \HST ($\lesssim0\farcs15$ in the near-IR) is able to resolve 
most galaxies at $z\leq3$. In addition, ancillary spectroscopic and imaging data 
combine with \HST data to provide tight constraints on galaxy redshifts, stellar masses, 
and star formation rates. CANDELS has three tiers of depth. The Wide region covers 
$\sim 675$ arcmin$^2$ to a $5\sigma$ limiting magnitude $H_{\rm 160} \sim 27.3$ mag in a
$0\farcs17$ aperture. The Deep region covers $\sim 125$ arcmin$^2$ to $H_{\rm 160} \sim 28.1$ mag.
The survey also encompasses the Hubble Ultra-Deep Field (HUDF)---the HUDF09
\citep{Bouwens:2010dk} and HUDF12 (\citealp{2013ApJ...763L...7E,2013ApJS..209....3K};
see also \citealp{2013ApJS..209....6I})---covers
$\sim 5$ arcmin$^2$ to $H_{\rm 160} \sim 29.7$ mag.

We take the photometry, spectroscopic and photometric redshifts, and stellar-mass estimates from
the CANDELS-team catalogs (\citealt{Guo:2013ig,Galametz:2013dd,Santini:2015hh,2016arXiv161207364N};
G. Barro et al. 2017, in preparation; M. Stefanon et al. 2017, in preparation). The size estimates
are taken from \cite{vanderWel:2012eu}.

We select galaxies in the CANDELS survey at $0 < z < 3$ for this study. We cap our
galaxy redshifts at $z=3$ because this is the highest redshift that \HST still
samples redward of rest-frame 4000\AA, and because selection biases induced by
cosmological surface brightness dimming are expected to be relatively mild for $z \leq 3$
\citep{TaghizadehPopp:2015hp}. Sources are detected using SExtractor \citep{Bertin:1996ww}
in $H_{160}$. Roughly 10\% of these sources have high-quality spectroscopic
redshifts, which are used in calibrating the photometric redshifts
for the remaining sources.


Galaxy sizes are measured in $H_{160}$ and $J_{125}$ 
by fitting a single S\'ersic profile to each galaxy using
GALFIT \citep{Peng:2010eh}. We define galaxy sizes as effective radii
($\Reff$) along the major axis, the radii within which S\'ersic profiles contain half of the total
integrated light. We discuss the deprojection from 2D to 3D later when
comparing with theoretical expectations. Our overall sample is dominated by late-type
galaxies at all redshifts, whose disk components
have the same 2D and 3D half-light radii.

Using simulations with artificial galaxies and comparisons of
measurements in different imaging depths, \cite{vanderWel:2012eu} concluded that
brighter than $H_{160}=24.5$ mag in the Wide region, the systematic (random)
errors of $\Reff$ measurements are below $\sim$20\% (30\%). Meanwhile, the
systematic (random) errors of S\'ersic index $n$ measurements are below
$\sim$50\% (60\%). The quoted errors here are for galaxies with $n>3$, which
tend to have larger errors than galaxies with $n<3$. Therefore, we select all
galaxies brighter than $H_{160}=24.5$ mag in the Wide region,
$H_{160}=25.2$ mag in the Deep region, and $H_{160}=26.7$ mag in the HUDF
(SExtractor-measured magnitudes). These
magnitude limits correspond to similar signal-to-noise limits.

In addition to magnitude cuts, we prune the sample as follows. 
We reject all sources that have problematic photometry (generally
those at the borders of the image or falling on stellar diffraction
spikes). We eliminate sources that are identified as active galactic nuclei
(AGNs) via X-ray or IR spectral energy distributions (SEDs). 
We discard as point sources all objects that have
half-light radii (measured by SExtractor) smaller than 2.6 pixels.
We enforce the following criteria to eliminate
galaxies with poor GALFIT fits: (1) the GALFIT measurement is
flagged as poor in
the catalogs from \cite{vanderWel:2012eu}; (2) the error in the measured $\Reff$
exceeds $0.3\Reff$; (3) the measured $n$ lies outside the range $0.1<n<8$,
which usually signals problematic fits. The GALFIT, AGN, and point-source
criteria combined reject roughly one-fourth of the sources that satisfy the magnitude cuts.
The numbers of sources that pass all the cuts above are listed in Table \ref{tab:sample}.

\begin{figure*}[ht]
\plotone{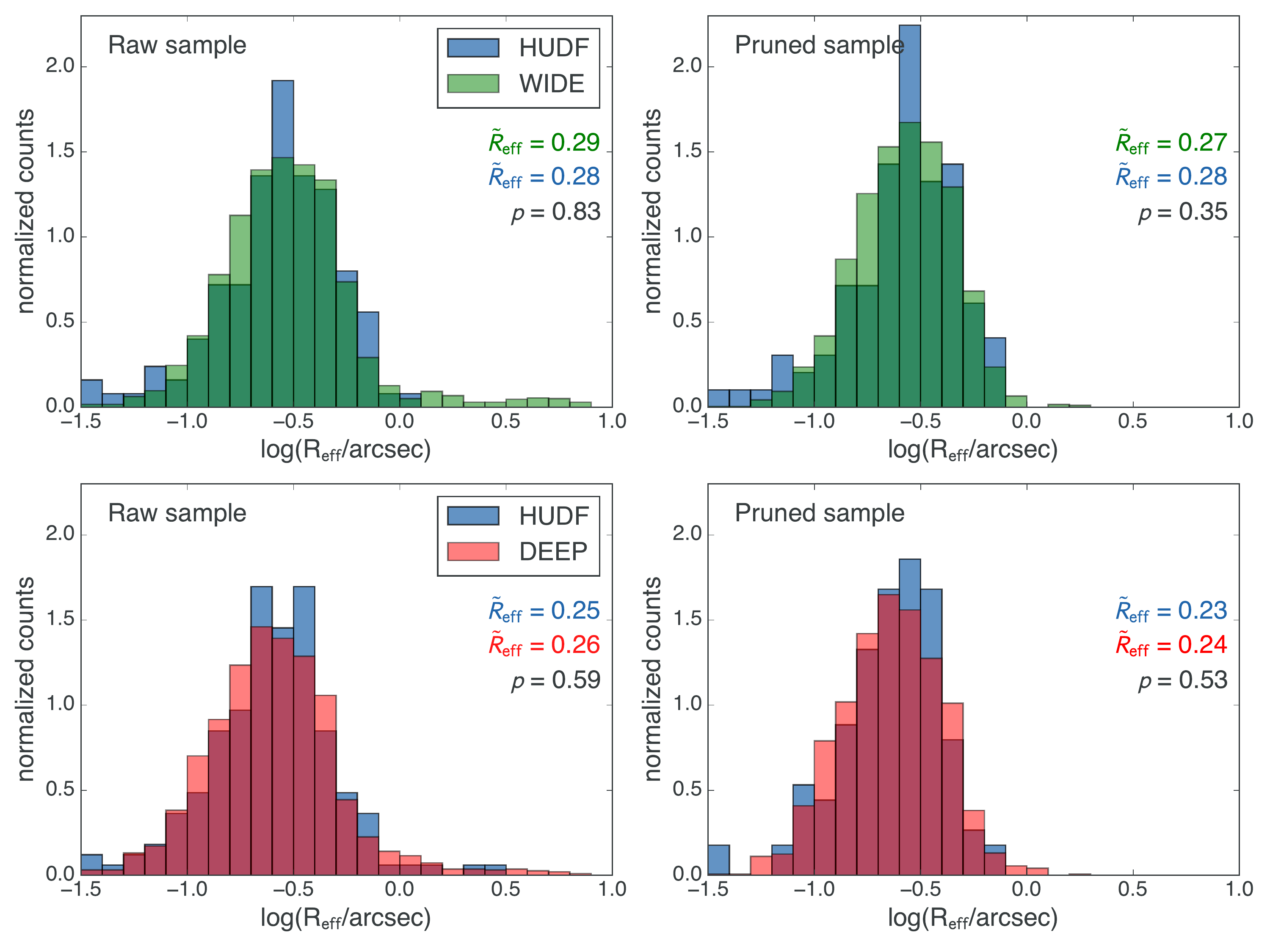}
\caption{
Histograms of effective radius $\Reff$ for galaxies in narrow magnitude ranges 
in the Wide, Deep, and HUDF regions of our sample.
The top panels compare the distributions of $\Reff$ in the Wide and HUDF
regions in the magnitude range $23.5$ mag $<H_{160}<24.5$ mag, while the bottom panels 
compare the distributions of $\Reff$ in the Deep and HUDF regions in the 
magnitude range $24.2$ mag $<H_{160}<25.2$ mag.
For reference, the selection limits of our sample in these regions are $H_{160} 
= 24.5$ (Wide), 25.2 (Deep), and 26.7 mag (HUDF).
The left and right panels compare the distributions before and after the sample
pruning described in Section \ref{sec:data}. 
The legends in the panels list the median values of $\Reff$ in the four histograms,
and the Kolmogorov-Smirnov probabilities that the histograms are drawn from the 
same underlying distribution.
The consistency of the histograms in regions with different depths,
before and after pruning, indicates that the distribution of galactic sizes in
our sample is unbiased even near the selection limits.
\label{fig:pruning}}
\end{figure*}

The existence of the very deep HUDF data allows us to test whether 
selection effects, measurement biases, or the pruning procedure are 
biasing our samples near their faint limits. In the top panels of Figure \ref{fig:pruning},
we compare the size distributions in the Wide region and the HUDF for
the magnitude range $23.5$ mag $<H_{\rm 160} < 24.5$ mag before and after pruning,
finding no significant difference. If the HUDF were picking up many 
more low surface brightness objects, we would have expected to see them
show up in the tail of the distribution. Instead, we see more large-radius
objects in the Wide sample, most of which are pruned away as bad fits,
but without having much impact on the median $\Reff$. A Kolmogorov-Smirnov test
yields $p$ values consistent with the samples being drawn from
the same underlying distribution. The bottom panels of Figure \ref{fig:pruning}
show the same comparison for the Deep region in the magnitude range
$24.2$ mag $<H_{160}<25.2$ mag.
We made a similar comparison for the stellar mass distributions, 
also finding no statistically significant difference between the HUDF 
and the Deep and Wide samples.

We have also estimated the completeness of our sample from the detection
efficiencies for the CANDELS survey derived by \cite{Guo:2013ig}. They inserted
artificial galaxies into images from the Wide, Deep, and HUDF regions and
analyzed them with SExtractor in the same way as the real survey to determine
the detection efficiency as a function of apparent magnitude $H_{160}$,
effective radius $\Reff$, and S\'ersic index $n$ (see their Fig. 5). From
these results, we estimate that our sample as a whole is more than 85\%
complete. This high level of completeness helps to ensure that selection
biases have relatively little impact on our galaxy size--halo size
relations (estimated in Section \ref{sec:errors}).

Studying galaxy size evolution demands that we compare $\Reff$ values at a similar
rest-frame wavelength across redshift bins, so that we can eliminate the
contributions from dust or stellar age gradient to the observed size evolution.
We follow the procedure in \cite{vanderWel:2014hi} to correct for galaxy color
gradients and place galaxy sizes on the same rest-frame wavelength. To do this,
we use galaxy sizes measured in $H_{160}$ for galaxies at $z>1.5$ and use the
sizes measured in $J_{125}$ at $z<1.5$. Color gradients that lead to different
galaxy sizes at different wavelengths are accounted for by a correction factor
that is a function of galaxy redshift, stellar mass, and galaxy type (late-type
or early-type). As the result of this color gradient correction, the measurements
are converted into the $\Reff$ near rest-frame 5000\AA. The size
correction is typically only a few percent, but it does reach $\sim$60\% in some cases.
For more details about the color gradient correction, we refer the readers to
\cite{vanderWel:2014hi}, Section 2.2, and their equations (1) and (2).

Stellar masses and star formation rates are estimated by comparing our photometry
with model SEDs, adopting a
\cite{Chabrier:2003ki} initial mass function (IMF).
Here the stellar masses of galaxies include all luminous stars
and dark remnants at the time of observation (but not stellar ejecta). 
This method of estimating
stellar masses has been extensively
tested in \cite{Mobasher:2015gp}, and they found that typical stellar mass
uncertainties are $\sim0.25$ dex for the magnitude limits adopted here. The
primary sources of systematic uncertainties are  IMF and stellar evolution
models; for galaxies with strong nebular emission lines, systematic
uncertainties for stellar mass can be up to $\sim 0.4$ dex.

We restrict this study to galaxies with stellar masses $M_* > 10^7\,M_{\odot}$.
Above this limit, we include all galaxies brighter than the magnitude limits mentioned
above, where we are confident that our measurements are robust and unaffected
by size-dependent biases. For each redshift interval, we estimate the typical
stellar mass of the faintest galaxies $M_{*,\rm{low}}$ by taking the median SED-fitted
stellar mass estimate of galaxies within 0.1 mag of the HUDF magnitude limit.
The values of $M_{*,\rm{low}}$ are listed in Table \ref{tab:sample} and shown as
thick tick marks at the bottoms of Figures \ref{fig:RR_z0}--\ref{fig:RR_allz_ssfr}. SED-based
star formation rates can be uncertain by $\sim0.4$ dex \citep{Salmon:2015iz}; therefore, the
uncertainties in the specific star formation rates (sSFRs) are roughly $\lesssim 0.6$ dex for
our galaxy sample. In this paper, we select subsamples in the upper and lower 20\%
tails of the sSFR distribution. Because we are making a differential
comparison between the relatively large populations in these tails,
our results are not sensitive to the sSFR uncertainties.


\section{Abundance Matching}\label{sec:matching}

In this study, we employ the technique of abundance matching to estimate the mass
and hence the size of the dark matter halo associated with each galaxy in our
sample. In essence, this technique compares the measured sizes of observed galaxies
with the inferred sizes of matched halos in cosmological dark matter simulations.
The basic assumption is that the rank ordering of galaxy (stellar) masses $\Mstellar$
reflects on average the rank ordering of halo (virial) masses $\Mhalo$, i.e.,
that the cumulative number densities of galaxy masses and halo masses are equal:
$n_g(>\Mstellar) = n_h(>\Mhalo)$. This ansatz leads directly to a correspondence
between $\Mstellar$ and $\Mhalo$ known as the stellar mass--halo mass
relation. While the assumption that galaxy masses and halo masses follow the same
rank ordering is a reasonable approximation for statistical studies based on large
samples such as ours, it cannot be exactly true for individual galaxies, which
experience stochastic events such as mergers and starbursts throughout their histories.

\begin{figure}[ht]
\plotone{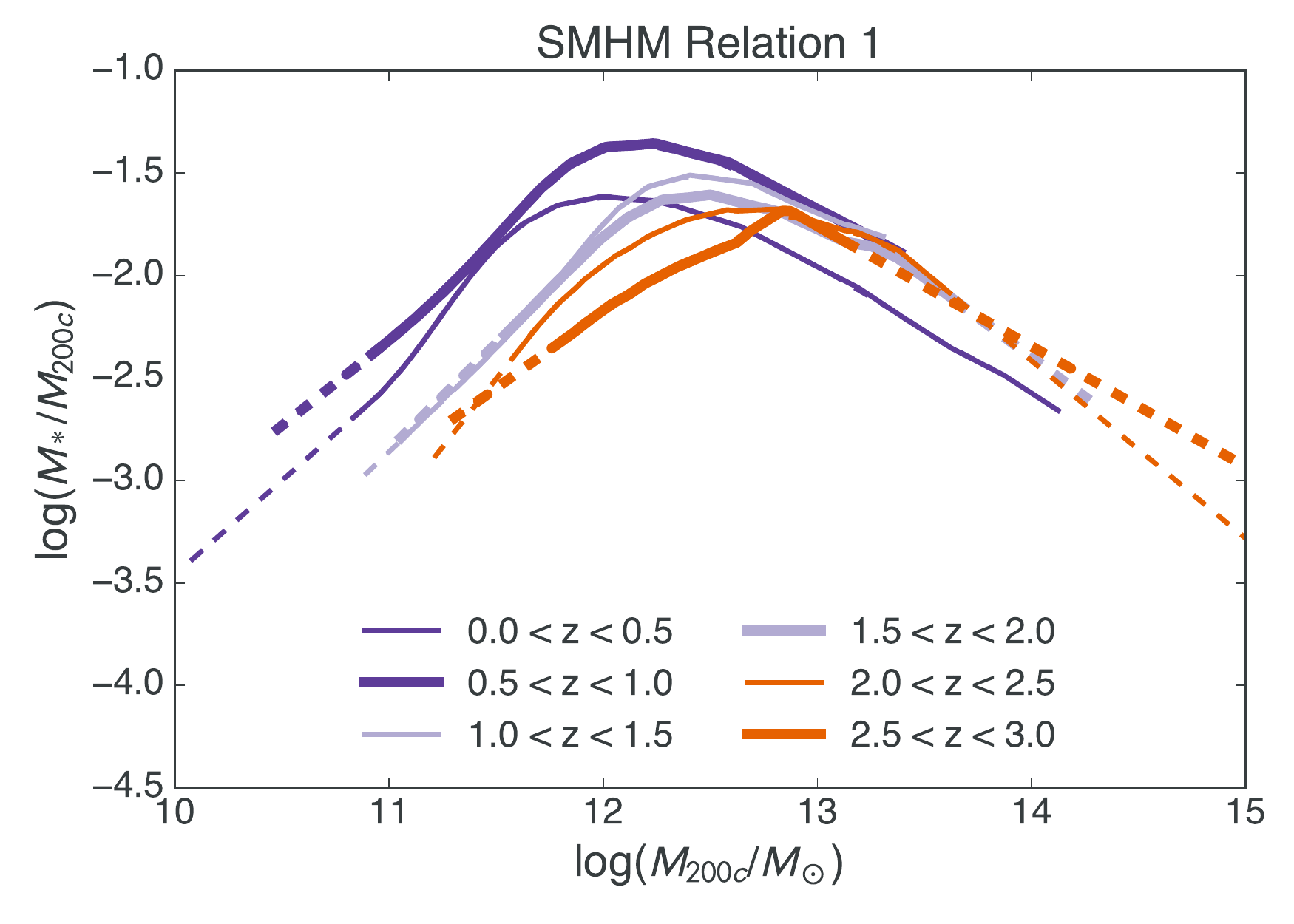}
\caption{Ratio of galaxy stellar mass $M_*$ to halo virial mass $\Mhalo$ plotted
against $\Mhalo$ for our primary SMHM relation in six redshift bins 
covering the
range $0 < z < 3$.  We derived this SMHM relation by abundance matching 
from an evolving stellar mass function appropriate for the CANDELS sample 
\citep{Tomczak:2014hw} and the evolving halo mass function in the 
Millennium-II simulation \citep{BoylanKolchin:2009co} as described in Section \ref{sec:matching}.
Solid lines are based directly on the stellar mass function
from \cite{Tomczak:2014hw}; we linearly extrapolate the SMHM relation in log--log space
to cover the stellar mass range of our sample (dashed lines).
\label{fig:SMHM1}}
\end{figure}
\begin{figure}[h]
\plotone{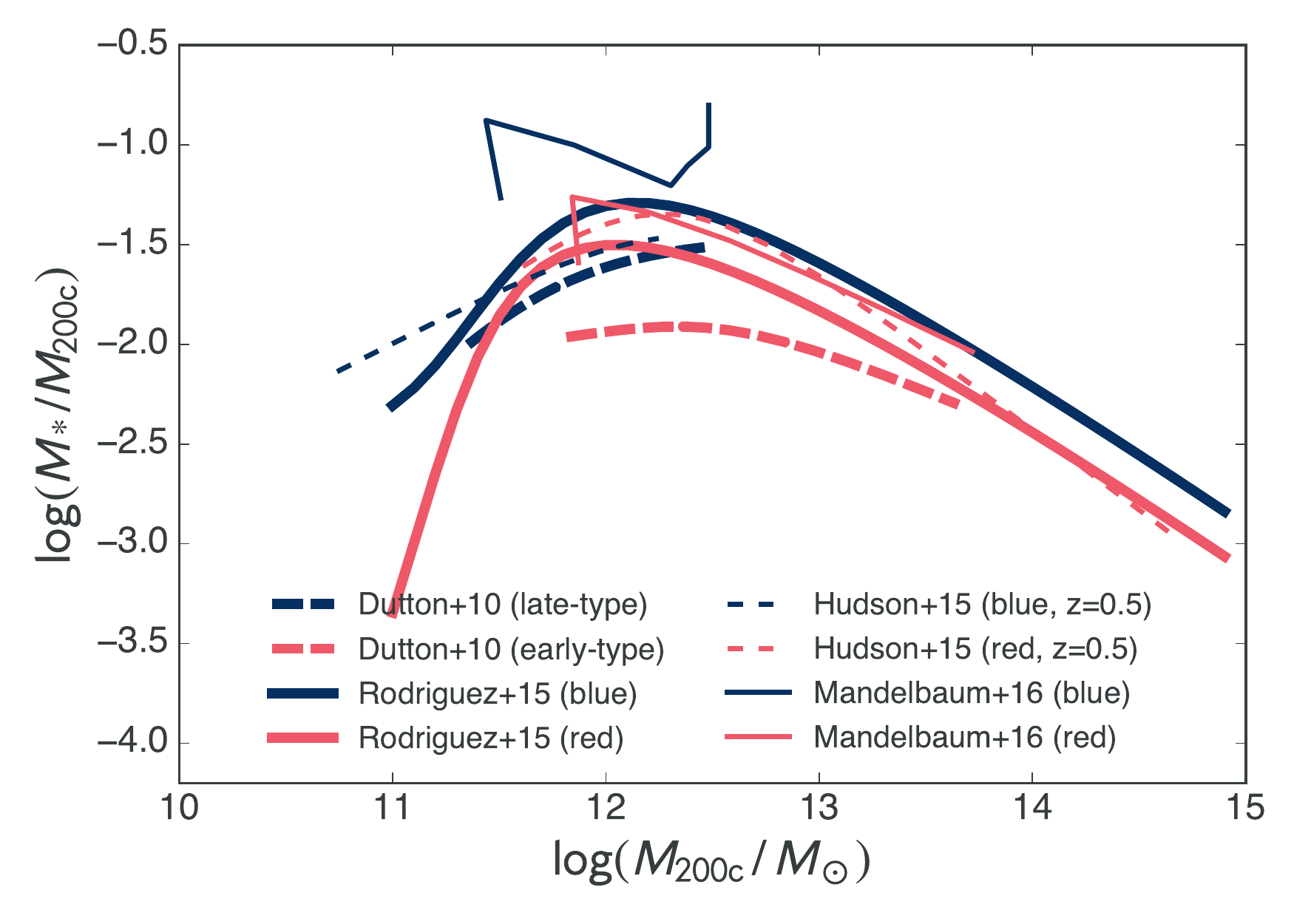}
\caption{Ratio of galaxy stellar mass $M_*$ to halo virial mass $\Mhalo$ plotted
against $\Mhalo$ for four low-redshift SMHM relations from the literature
that depend on galaxy color or type. These were derived by abundance 
matching \citep{RodriguezPuebla:2015bk}, weak lensing 
\citep{2015MNRAS.447..298H, Mandelbaum:2016eb}, 
or a combination of both techniques \citep{Dutton:2010hf}. 
Three of the SMHM relations pertain to $z = 0$ and one to
$z = 0.5$ \citep{2015MNRAS.447..298H}. 
Note the large discrepancies among these color- and 
type-dependent SMHM relations. \label{fig:SMHM_gtypes}}
\end{figure}
\begin{figure}[h]
\plotone{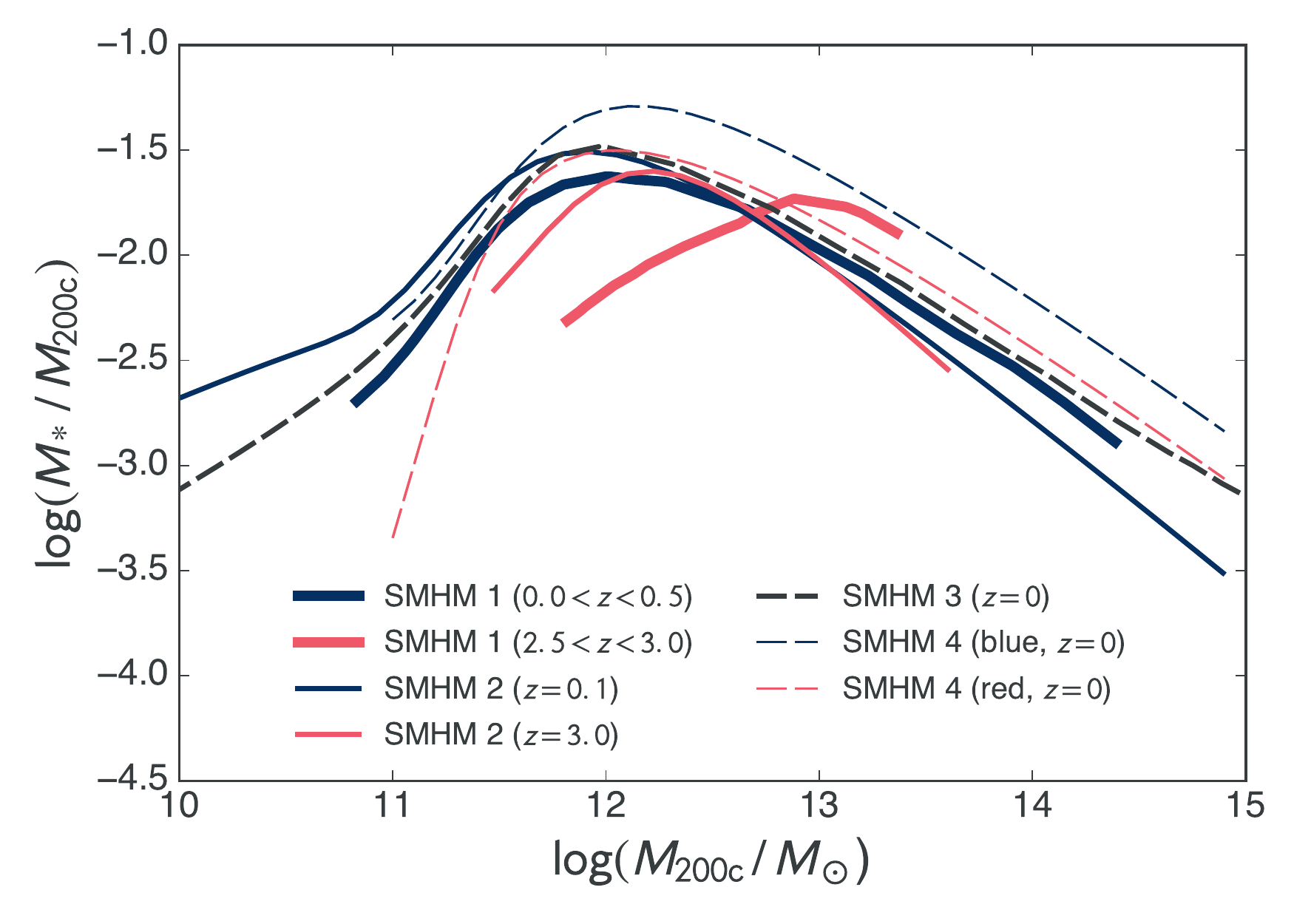}
\caption{Ratio of galaxy stellar mass $M_*$ to halo virial mass $\Mhalo$ plotted
against $\Mhalo$ for the four SMHM relations adopted in this work.
\emph{SMHM relation 1}: derived as described in Section 3 for all
galaxies at $0< z <3$ and displayed here at $0 < z < 0.5$ and 
$2.5 < z < 3.0$, which bracket the relation at intermediate redshifts. 
\emph{SMHM relation 2}: derived by \cite{Behroozi:2013fg} for all
galaxies at $0<z<8$ and displayed here at $z = 0.1$ and $z = 3.0$.
\emph{SMHM relation 3}: derived by \cite{Kravtsov:2013cy} for all
galaxies only at $z=0$.
\emph{SMHM relation 4}: derived by \cite{RodriguezPuebla:2015bk} 
separately for blue and red galaxies only at $z = 0$.
Note that there are significant differences among these SMHM 
relations, but because halo size depends weakly on halo mass
($\Rhalo \propto \Mhalo^{1/3}$), our main results are not sensitive to these 
differences.\label{fig:smhm}}
\end{figure}

Given an SMHM relation, we compute the halo mass $\Mhalo$ of each galaxy
in our sample from its stellar mass $\Mstellar$. We then compute the virial 
halo radius $\Rhalo$ using the standard formula
\begin{equation}\label{eq:rvir}
\Rhalo = \left[\frac{3\Mhalo}{4\pi\cdot200\rho_{\rm{crit}}(z)}\right]^{1/3},
\end{equation}
where $\rho_{\rm{crit}}(z)$ is the critical density of the universe at redshift $z$.
In order to 
assess how sensitive our results are to the choice of SMHM relation, we perform all 
of our calculations with four different SMHM relations. All of these SMHM relations are
based on the \cite{Chabrier:2003ki} stellar IMF and 
the same halo mass definition ${\Mhalo}$. They are plotted in Figures
\ref{fig:SMHM1}, \ref{fig:SMHM_gtypes}, and \ref{fig:smhm} and discussed below.

\emph{SMHM relation 1}. We have derived this new SMHM relation specifically for
this study so that it is as consistent as possible with the CANDELS data set, selection
criteria, and SED fitting procedure for our sample of galaxies with size measurements.
In particular, we combine the stellar mass function $n_g(>\Mstellar)$ 
from \cite{Tomczak:2014hw} with our
determination of the halo mass function $n_h(>\Mhalo)$ from the Millennium-II
simulation \citep{BoylanKolchin:2009co}.


\cite{Tomczak:2014hw} derived the stellar mass function of galaxies at $0.2 < z < 3$
in three of the five CANDELS fields, using selection criteria and
procedures for estimating stellar masses similar to those for our sample,
as described in Section \ref{sec:data}.
We have compared our stellar masses with those derived by \cite{Tomczak:2014hw}
\footnote{These stellar masses are published by the ZFOURGE team \citep{2016ApJ...830...51S}
and can be downloaded from \url{http://zfourge.tamu.edu}.} 
and find no systematic offset and only a small scatter ($\sim 0.1$ dex).
\citeauthor{Tomczak:2014hw} fitted a double Schechter function to the
observed stellar mass function in
differential form $dn_g(>\Mstellar)/d\Mstellar$ in each of eight redshift bins. We adopt the
\citeauthor{Tomczak:2014hw} results directly for the three bins of width
$\Delta z=0.5$ covering the range $1.5 < z < 3.0$. However, for simplicity, we
combine their results for the four bins of width $\Delta z=0.25$ covering the range
$0.5 < z < 1.5$ into two bins of width $\Delta z=0.5$. In this step, we weight the
observed comoving densities of galaxies by the comoving volume in each $\Delta z=0.25$
bin and then fit a double Schechter function to the combined comoving densities in each
$\Delta z=0.5$ bin. For our lowest redshift bin, $0 < z < 0.5$,
we adopt the \citeauthor{Tomczak:2014hw}
stellar mass function in their lowest redshift bin, $0.2 < z < 0.5$, because it
agrees well with the one at $<z>=0.1$ derived by \cite{Moustakas:2013il}.
Finally, we have derived the halo mass function $n_h(>\Mhalo)$ from the
Millennium-II simulation \citep{BoylanKolchin:2009co} at the snapshot closest
to the middle of each redshift bin and then matched this to the stellar mass function
as described above to obtain the SMHM relation.

As a check on this procedure, we have independently derived our own stellar mass
function from scratch by the $1/V_{\rm{max}}$ method for the galaxies in all five
CANDELS fields in the six $\Delta z=0.5$ bins (albeit with approximate
$K$-corrections in our estimates of $V_{\rm{max}}$). The resulting stellar mass function 
is nearly identical to the rebinned one from \cite{Tomczak:2014hw}. This adds to 
our confidence in the validity of SMHM relation 1,
which we regard as the primary SMHM relation in this study.

Because our galaxy sample covers a wider range in stellar mass than the
\citeauthor{Tomczak:2014hw} sample, we linearly extrapolate the SMHM relation in log--log space
to both lower and higher masses. The solid lines in Figure \ref{fig:SMHM1}
show the SMHM relation derived directly from the \citeauthor{Tomczak:2014hw} data,
while the dashed lines show the extrapolated parts of the SMHM relation.



\emph{SMHM relation 2}. \cite{Behroozi:2013fg} derived this SMHM relation from published
stellar mass and halo mass functions over a wide range of redshifts ($0 < z < 8$).
This is probably the most prevalent
SMHM relation in the literature. However, since it is based on stellar mass functions
that are quite different from those derived using CANDELS data, it is not ideal
for the present study. We use it mainly to gauge the sensitivity of our results to
different SMHM relations.
For consistency, we convert their halo mass $M_{\rm{vir}}$, defined using a 
redshift-dependent overdensity factor $\Delta_{\rm{vir}}(z)$ \citep{Bryan:1998cc},
to our halo mass definition
$\Mhalo$. The conversion assumes an NFW halo mass profile and the halo mass--concentration
model calibrated in \cite{Diemer:2015bd}. The corrections are very small
in general ($< 0.1$ dex).


\emph{SMHM relation 3}. This is the same SMHM relation adopted by \cite{Kravtsov:2013cy}. 
He derived his own SMHM relation out of concerns that previous relations used stellar 
mass functions that are biased at both the high-mass and low-mass ends. 
By using the same SMHM relation as \cite{Kravtsov:2013cy}, we can directly compare our 
galaxy size--halo size relation with his at $z=0$.

\emph{SMHM relation 4}. There are several SMHM relations separated by
galaxy type at $z < 0.5$ in the literature, which we plot 
in Figure \ref{fig:SMHM_gtypes}. These relations use different approaches
to deriving the ratio between stellar masses and
halo masses, ranging from abundance matching \citep{RodriguezPuebla:2015bk} to 
weak lensing \citep{2015MNRAS.447..298H, Mandelbaum:2016eb} to a mixture of 
the two methods \citep{Dutton:2010hf}. We adopt the SMHM relation from 
\cite{RodriguezPuebla:2015bk}
because it has the largest dynamic range in halo mass and is in the middle of 
the range spanned by the other type-dependent relations from the literature.
We use the \citeauthor{RodriguezPuebla:2015bk} SMHM relations for blue and
red central galaxies at $z=0$ for galaxies in our sample with S\'ersic index
$n$ below and above 2.5, respectively.
Since \citeauthor{RodriguezPuebla:2015bk} defined their halo mass using
$\Delta_{\rm{vir}}(z)$, we have applied the same conversion to $\Mhalo$ as we did
for SMHM relation 2.

We compare the four SMHM relations in Figure \ref{fig:smhm}. Evidently,
there are significant discrepancies among these SMHM relations, especially
the first and second, for which the differences can be up to $\sim 0.5$ 
dex at $z \sim 3$. Our SMHM relation 1, derived specifically for the CANDELS sample 
at $0 < z < 3$, shows stronger redshift evolution than SMHM relation 2 from
\cite{Behroozi:2013fg}. As already noted, this difference comes mainly from the different 
stellar mass functions used as input to these SMHM relations. Fortunately, as we
show in Sections \ref{sec:results} and \ref{sec:errors}, our main scientific results
are relatively insensitive to the adopted SMHM relation, largely due to the weak
dependence of halo size on halo mass ($\Rhalo \propto \Mhalo^{1/3}$).

\section{Results}\label{sec:results}

\begin{figure*}[ht]
\begin{center}
\plotone{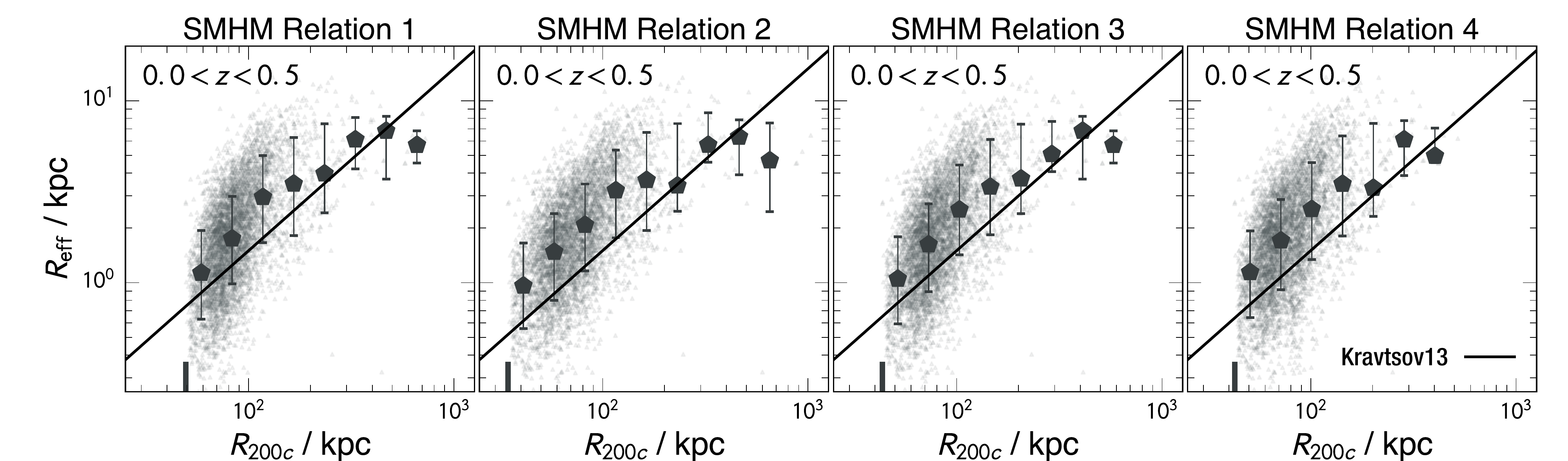}
\figcaption{Galaxy effective radius $\Reff$ plotted against halo virial radius $\Rhalo$ in the 
lowest redshift interval ($0 < z < 0.5$) for the full sample of galaxies.
The four panels show results for SMHM relations 1, 2, 3, and 4 as indicated.
The faint gray dots represent individual galaxies, while the filled pentagons and vertical
bars indicate the median values and 16th--84th percentile ranges of $\Reff$ in bins of
width 0.15 in $\log\Rhalo$.
The diagonal lines show the $R_{1/2}$--$\Rhalo$ relation at $z=0$ from \cite{Kravtsov:2013cy} 
assuming $\Reff=R_{1/2}$.
The thick tick mark at the bottom of each panel indicates the halo size corresponding
to the reference stellar mass $M_{*,\rm{low}}$ listed in Table \ref{tab:sample}.
Note that the $\Reff$--$\Rhalo$ relations are similar for the four different SMHM relations
and are roughly consistent with Kravtsov's results. 
The $\Reff$--$\Rhalo$ relations are linear in a first approximation but exhibit some 
curvature at high and low masses as a result of the changing mix of galaxy morphologies. 
Compare with Figures \ref{fig:RR_z0_sersic} and \ref{fig:RR_z0_ssfr}.
\label{fig:RR_z0}}
\end{center}
\end{figure*}

\begin{figure*}[ht]
\begin{center}
\plotone{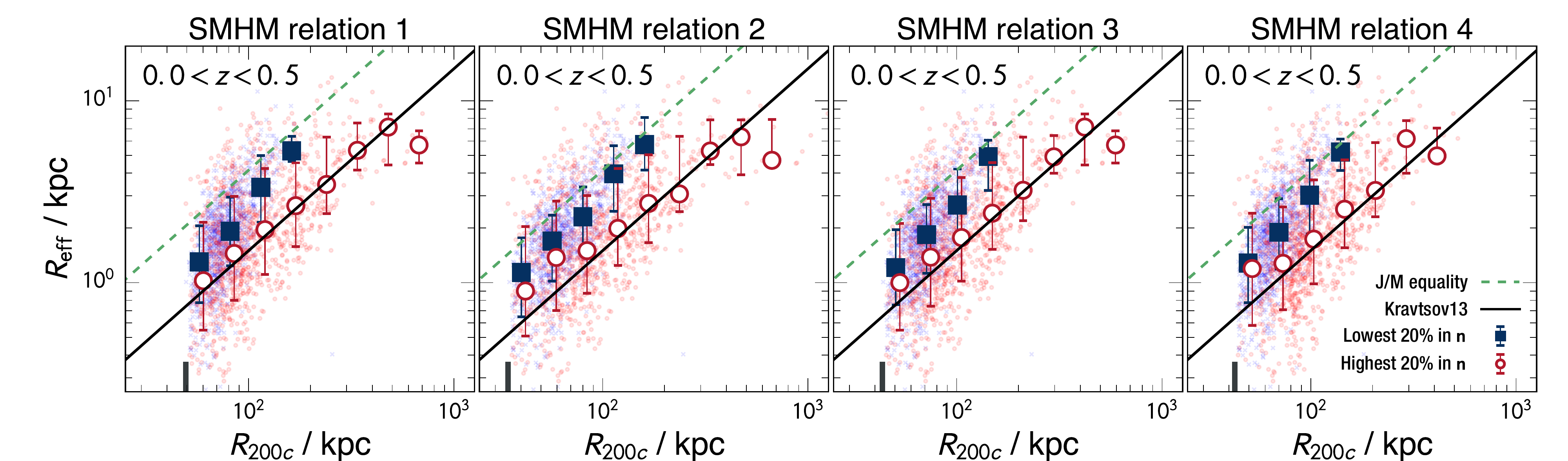}
\figcaption{Galaxy effective radius $\Reff$ plotted against halo virial radius $\Rhalo$ in the 
lowest redshift interval ($0 < z < 0.5$) for subsamples of galaxies with the lowest
and highest 20\% of the measured S\'ersic index $n$ as proxies for late- 
and early-type galaxies, respectively.
The four panels show results for SMHM relations 1, 2, 3, and 4 as indicated.
The faint blue and red dots represent individual low-$n$ and high-$n$ galaxies, 
respectively, while the filled blue squares, open red circles, and vertical bars 
indicate the corresponding median values and 16th--84th percentile ranges of $\Reff$ 
in bins of width 0.15 in $\log\Rhalo$.
The diagonal solid lines show the $R_{1/2}$--$\Rhalo$ relation at $z=0$ from 
\cite{Kravtsov:2013cy} assuming $\Reff=R_{1/2}$, while the diagonal dashed 
lines show the prediction for galactic disks with the same $J/M$ as
their surrounding halos.
The thick tick mark at the bottom of each panel indicates the halo size corresponding
to the reference stellar mass $M_{*,\rm{low}}$ listed in Table \ref{tab:sample}.
Note that the $\Reff$--$\Rhalo$ relation for low-$n$ galaxies is systematically 
above, and roughly parallel to, the relation for high-$n$ galaxies.
The $\Reff$--$\Rhalo$ relations for both subsamples of galaxies are more linear
than the relations for the full sample.  Compare with Figures \ref{fig:RR_z0} and \ref{fig:RR_z0_ssfr}.
\label{fig:RR_z0_sersic}}
\end{center}
\end{figure*}

\begin{figure*}[ht]
\begin{center}
\plotone{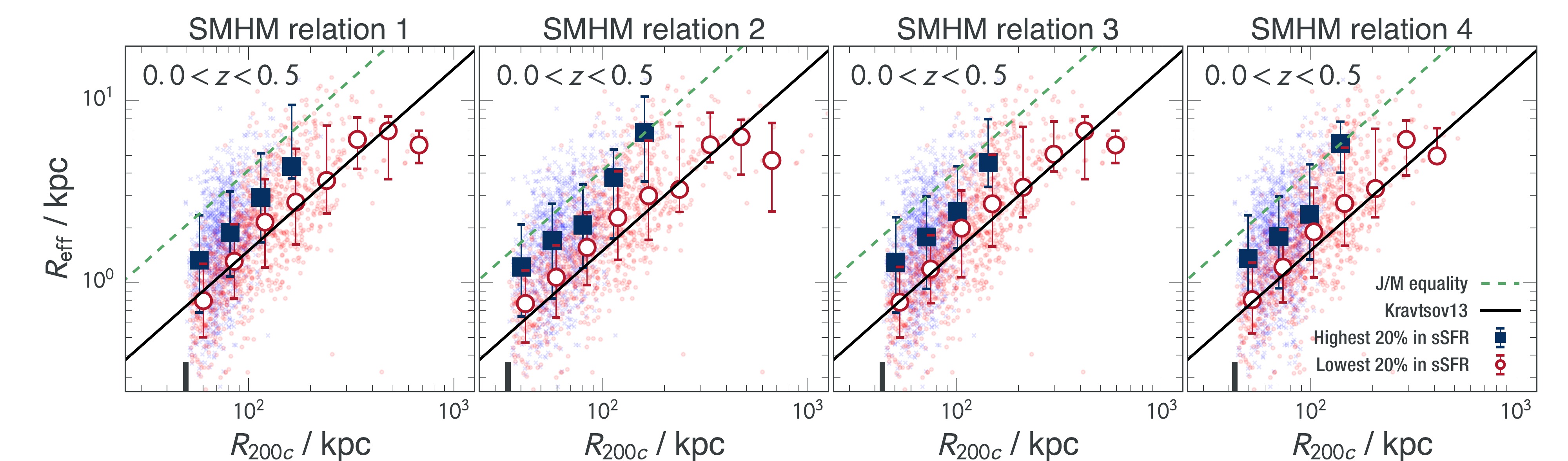}
\figcaption{Galaxy effective radius $\Reff$ plotted against halo virial radius $\Rhalo$ in the 
lowest redshift interval ($0 < z < 0.5$) for subsamples of galaxies with the highest
and lowest 20\% of the measured sSFR as proxies 
for late- and early-type galaxies, respectively.
The four panels show results for SMHM relations 1, 2, 3, and 4 as indicated.
The faint blue and red dots represent individual high-sSFR and low-sSFR galaxies, 
respectively, while the filled blue squares, open red circles, and vertical bars 
indicate the corresponding median values and 16th--84th percentile ranges of $\Reff$ 
in bins of width 0.15 in $\log\Rhalo$.
The diagonal solid lines show the $R_{1/2}$--$\Rhalo$ relation at $z=0$ from 
\cite{Kravtsov:2013cy} assuming $\Reff=R_{1/2}$, while the diagonal dashed 
lines show the prediction for galactic disks with the same $J/M$ as
their surrounding halos.
The thick tick mark at the bottom of each panel indicates the halo size corresponding
to the reference stellar mass $M_{*,\rm{low}}$ listed in Table \ref{tab:sample}.
Note that the $\Reff$--$\Rhalo$ relation for high-sSFR galaxies is systematically 
above, and roughly parallel to, the relation for low-sSFR galaxies.
The $\Reff$--$\Rhalo$ relations for both subsamples of galaxies are more linear
than the relations for the full sample.  Compare with Figures \ref{fig:RR_z0} and \ref{fig:RR_z0_sersic}.
\label{fig:RR_z0_ssfr}}
\end{center}
\end{figure*}

\begin{figure*}[ht]
\begin{center}
\plotone{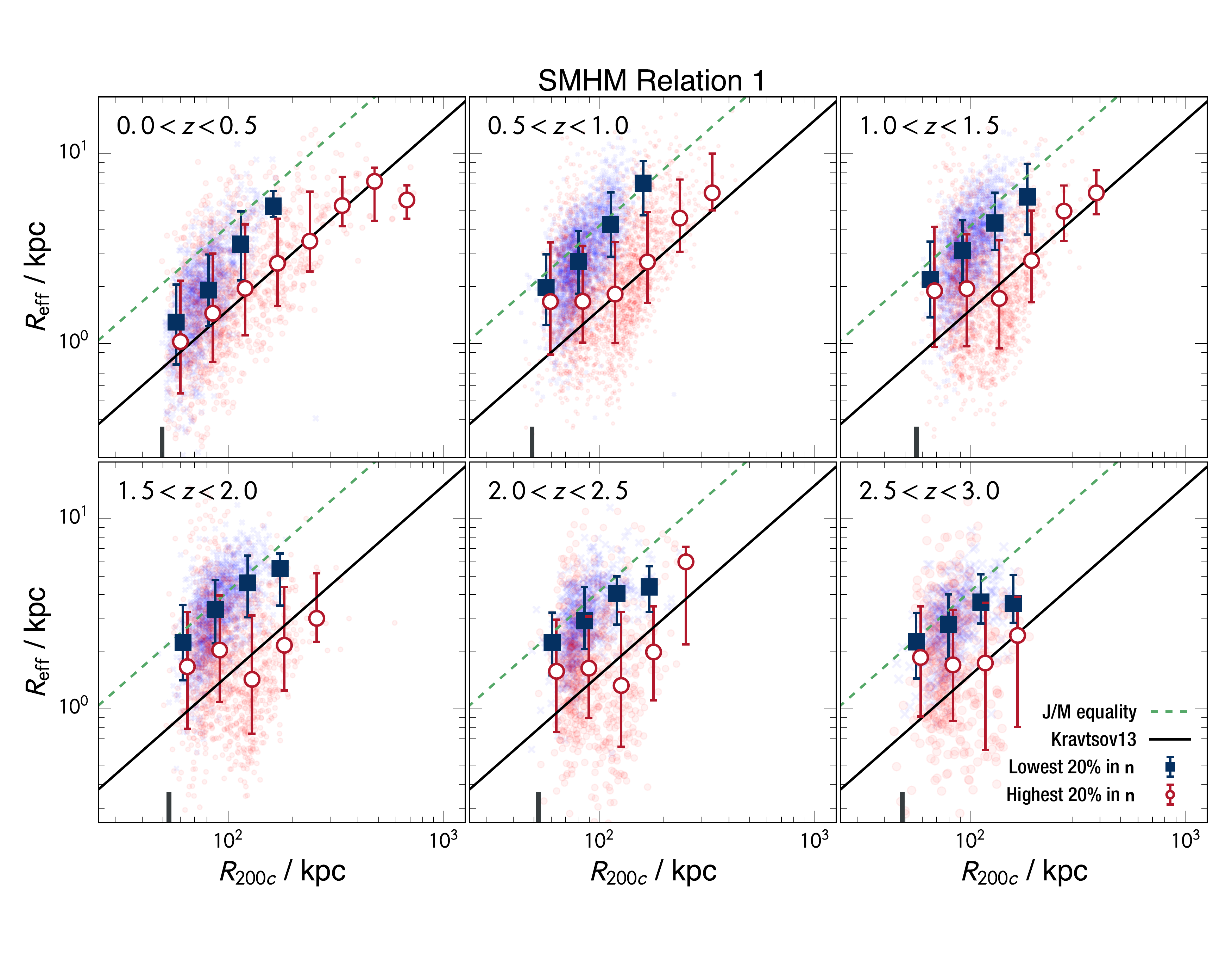}
\figcaption{Galaxy effective radius $\Reff$ plotted against halo virial radius $\Rhalo$ at
different redshifts for subsamples of galaxies with the lowest and highest 20\% 
of the measured S\'ersic index $n$ as proxies for late- and early-type 
galaxies, respectively.
The six panels show results computed from SMHM relation 1
in redshift intervals of $\Delta z = 0.5$ covering the range $0 < z < 3$.
The faint blue and red dots represent individual low-$n$ and high-$n$ galaxies, 
respectively, while the filled blue squares, open red circles, and vertical bars 
indicate the corresponding median values and 16th--84th percentile ranges of $\Reff$ 
in bins of width 0.15 in $\log\Rhalo$.
The diagonal solid lines show the $R_{1/2}$--$\Rhalo$ relation at $z=0$ from 
\cite{Kravtsov:2013cy} assuming $\Reff=R_{1/2}$, while the diagonal dashed 
lines show the prediction for galactic disks with the same $J/M$ as
their surrounding halos.
The thick tick mark at the bottom of each panel indicates the halo size corresponding
to the reference stellar mass $M_{*,\rm{low}}$ listed in Table \ref{tab:sample}.
Note that the $\Reff$--$\Rhalo$ relations for both low-$n$ and high-$n$ galaxies are
nearly constant with redshift, and that the one for low-$n$ galaxies is close to the 
predicted relation for equality of $J/M$ in disks and halos. Compare with Figure \ref{fig:RR_allz_ssfr}.
\label{fig:RR_allz_sersic}}
\end{center}
\end{figure*}

\begin{figure*}
\begin{center}
\plotone{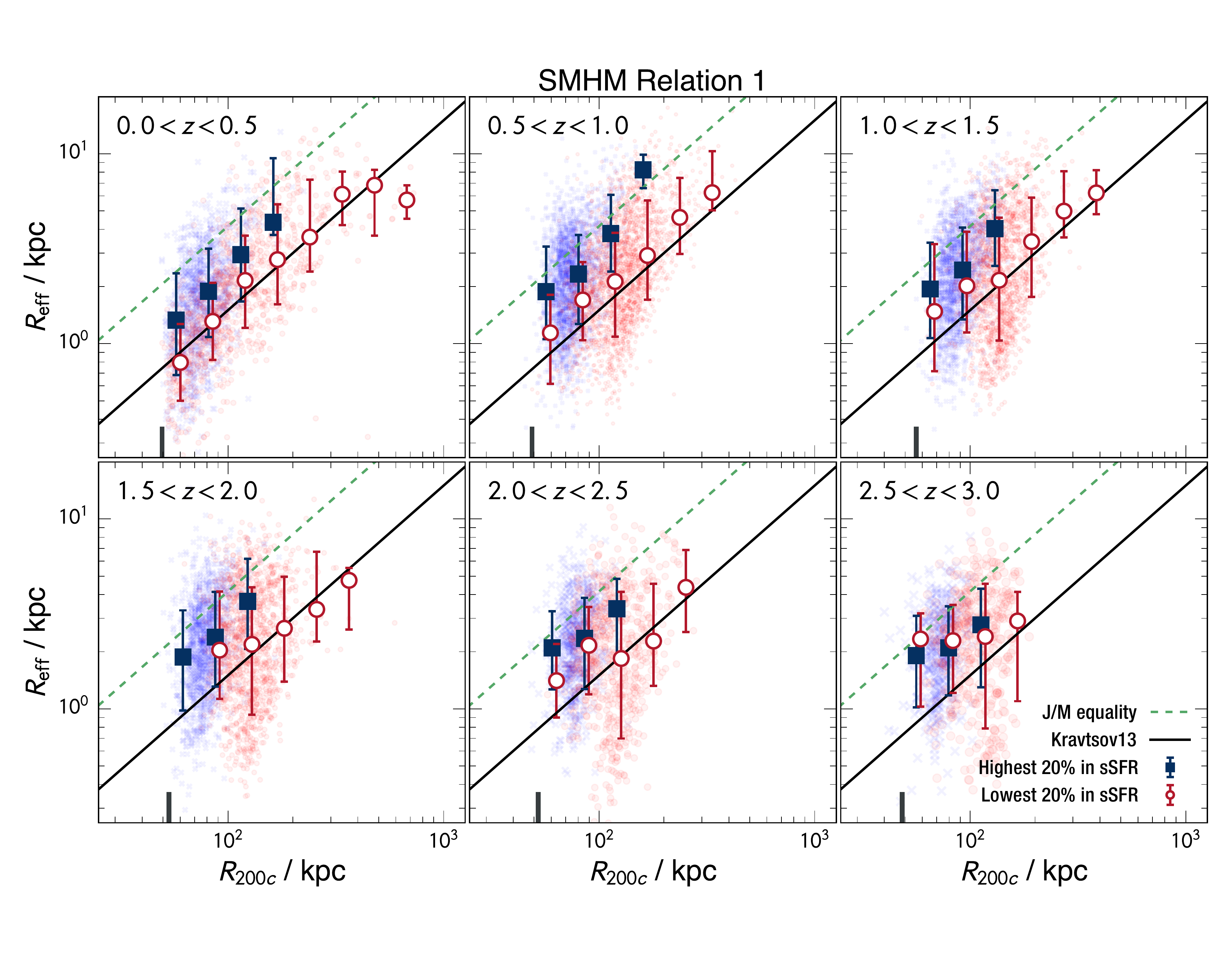}
\figcaption{Galaxy effective radius $\Reff$ plotted against halo virial radius $\Rhalo$ at
different redshifts for subsamples of galaxies with the highest and lowest 20\% 
of the measured sSFR as proxies for late- and early-type 
galaxies, respectively.
The six panels show results computed from SMHM relation 1
in redshift intervals of $\Delta z = 0.5$ covering the range $0 < z < 3$.
The faint blue and red dots represent individual high-sSFR and low-sSFR galaxies, 
respectively, while the filled blue squares, open red circles, and vertical bars 
indicate the corresponding median values and 16th--84th percentile ranges of $\Reff$ 
in bins of width 0.15 in $\log\Rhalo$.
The diagonal solid lines show the $R_{1/2}$--$\Rhalo$ relation at $z=0$ from 
\cite{Kravtsov:2013cy} assuming $\Reff=R_{1/2}$, while the diagonal dashed 
lines show the prediction for galactic disks with the same $J/M$ as
their surrounding halos.
The thick tick mark at the bottom of each panel indicates the halo size corresponding
to the reference stellar mass $M_{*,\rm{low}}$ listed in Table \ref{tab:sample}.
Note that the $\Reff$--$\Rhalo$ relations for both high-sSFR and low-sSFR galaxies are
nearly constant with redshift, and that the one for high-sSFR galaxies is close to the 
predicted relation for equality of $J/M$ in disks and halos. Compare with Figure \ref{fig:RR_allz_sersic}.
\label{fig:RR_allz_ssfr}}
\end{center}
\end{figure*}

The main results of this paper are displayed in Figures
\ref{fig:RR_z0}--\ref{fig:RR_allz_ssfr} and described in this section. The uncertainties 
in these results, mostly stemming from the SMHM relation and morphological 
classification, are discussed in Section \ref{sec:errors}.

Our first main result is that galaxy sizes are proportional to halo sizes
over a wide range of size and mass. Figure \ref{fig:RR_z0} shows galaxy
$\Reff$ plotted against halo $\Rhalo$ at $0 < z < 0.5$ for the four different SMHM relations. 
In each panel, the medians of
$\log\Reff$ in bins of width $\Delta\log\Rhalo=0.15$ dex are plotted as pentagons,
and the 16th--84th percentile ranges as vertical bars; only the bins with 
more than five galaxies are shown. The halo
radius limit corresponding to the reference stellar mass $M_{*,\rm{low}}$
from Table \ref{tab:sample}
is shown as a thick tick mark at the bottom of each panel.
The coefficient of
proportionality $\alpha$ in the relation $\Reff=\alpha\Rhalo$ is nearly the same
in all four cases; the median values of $\alpha$ are 0.021, 0.025, 0.023,
and 0.024 for SMHM relations 1--4, respectively. These $\Reff$--$\Rhalo$ relations
are approximately linear, but with some subtle differences depending on the adopted
SMHM relation.

\citet{Kravtsov:2013cy} also found a linear relation, using completely independent
samples of galaxies at $z = 0$ and de-projected 3D half-\emph{mass}
radii $R_{1/2}$ rather than the projected 2D half-\emph{light} radii $\Reff$. The solid line 
in Figure \ref{fig:RR_z0} shows his 
derived relation $R_{1/2}=\alpha^\prime\Rhalo$ with $\alpha^\prime=0.015$, assuming $\Reff=R_{1/2}$ for 
pure-disk galaxies. The bulk
of our sample by number lies above this relation by ${\sim0.2}$ dex, agreeing better 
at the high- and low-mass ends. There are a number of possible explanations for this offset,
one of them being the difference between 2D half-light (effective) and 2D half-mass radii.
\cite{Szomoru:2013gh} noted that for 
the galaxies more massive than $5\times10^{10}~M_\odot$ at $0 < z < 2.5$, rest-frame $g$-band
2D half-light radii are on average $\sim$25\% larger than 2D half-mass radii
(presumably due to the influence of bulges), which could account for $\sim 0.1$ dex of the offset.
We will address other explanations below in connection with morphological
types, deprojection effects, and the redshift evolution. 

Our second main result is that the $\Reff$--$\Rhalo$ relations are offset for 
late-type and early-type galaxies.  
To separate morphological types, we split our
sample in two different ways: (1) high-$n$ (early-type)
and low-$n$ (late-type) subsamples, and (2) low-sSFR
(early-type) and high-sSFR (late-type) subsamples. We only include the highest
and lowest 20\% of the sample in either $n$ or sSFR in the hope that
this procedure will isolate disk-dominated from spheroid-dominated
galaxies. The resulting
$\Reff$--$\Rhalo$ relations for late- and early-type galaxies using all four
SMHM relations are shown in Figures \ref{fig:RR_z0_sersic} and
\ref{fig:RR_z0_ssfr}.

We see in both Figures \ref{fig:RR_z0_sersic} and \ref{fig:RR_z0_ssfr} that
galaxies of different types follow sequences roughly parallel to the 
$\Reff\propto\Rhalo$ line with
an offset of $\sim 0.2$ dex at $0 < z < 0.5$. This result is relatively robust against
SMHM relation and morphological classification method: early-type
(high-$n$ or low-sSFR) galaxies have smaller $\Reff$ than late-type (low-$n$ or
high-sSFR) galaxies at the same halo masses. The effect persists even if we
compare 3D half-light radii rather than 2D half-light radii $\Reff$, although
with a smaller separation between the sequences. The parallel sequences of early-
and late-type galaxies in the $\Reff$--$\Rhalo$ diagram are reminiscent of the
parallel sequences of spheroid- and disk-dominated galaxies in the $J/M$ vs. $M$
diagram \citep{Fall:1983wu,Romanowsky:2012kb,Fall:2013du}. The latter is due to
a combination of different sizes (by a factor of $\sim$2) and different 
rotation velocities (also by a factor of $\sim$2--3) of spheroid- and disk-dominated
galaxies of the same stellar mass.


This helps explain why our overall relation in Figure \ref{fig:RR_z0} is higher than
Kravtsov's at intermediate masses. 
Our sample is dominated by late-type galaxies
($\sim$90\% have $n<2.5$), while Kravtsov's sample is dominated
by early-type galaxies ($\sim 80\%$ by number). 
He noted that late-type galaxies are systematically larger in $R_{1/2}$ than early-type galaxies at intermediate
stellar masses, which is where we see the largest offset between these sequences in
Figure \ref{fig:RR_z0}.
The changing morphological
mix as a function of mass also helps explain the apparent curvature of the
overall relation in Figure \ref{fig:RR_z0}, because early-type galaxies 
dominate the high- and low-mass ends of the relation.


Our third main result is that the $\Reff$--$\Rhalo$ relation for
late-type galaxies is close to the predictions of the simple analytic model of
disk formation. The scale radius and
effective radius of an exponential disk embedded in a dark matter halo
with a virial (outer) radius $\Rhalo$ and a spin parameter $\lambda$ are given by
\begin{equation}\label{eq:RR_disk}
R_d = \frac{\lambda}{\sqrt{2}}\Rhalo
\end{equation}
and
\begin{equation}\label{eq:RR_eff}
\Reff = 1.68~R_d,
\end{equation}
when the disk and halo have the same specific angular momentum ($J/M$). Equation
(\ref{eq:RR_disk}) is exact for isothermal halos (\citealp{Fall:1980up}; see their 
Figure 3 and equation 42; \citealp{Fall:1983wu}, see his equation 4) and is
approximate for NFW halos with typical concentrations \citep{Mo:1998hg,Burkert:2016fr}. 
This prediction is shown as the dashed lines in Figures \ref{fig:RR_z0_sersic} to
\ref{fig:RR_allz_ssfr} for $\lambda=0.035$, the peak of the universal spin
parameter distribution \citep{Bullock:2001kb,2007MNRAS.376..215B}. We find that
late-type galaxies at $0 < z < 0.5$ lie $\sim 0.2$ dex below the $J/M$
equality line; in other words, our late-type galaxies have slightly less specific
angular momentum than their dark matter halos. This offset is consistent with direct
measurements of specific angular momentum at $z=0$, which indicate $J/M$ retention factors
$\eta_j \sim 80\%\pm20\%$ for galactic disks \citep{Fall:2013du}. 


Our fourth main result is that there is remarkably little evolution in the 
$\Reff$--$\Rhalo$ relation from $z=3$ to $z=0$.  This is shown 
in Figures \ref{fig:RR_allz_sersic} and \ref{fig:RR_allz_ssfr}. 
As in the previous diagrams, we select the highest
and lowest 20\% tails of the $n$ and sSFR distributions. We only show
results for SMHM relation 1, but we have checked that they
are similar for the other SMHM relations. Figures
\ref{fig:RR_allz_sersic} and \ref{fig:RR_allz_ssfr} show again that in all
redshift bins, late-type galaxies follow a nearly linear relation: $\Reff=\alpha\Rhalo$.
At $0.5 < z < 3$, late-type galaxies have ${\alpha \approx 0.034}$ in
Figure \ref{fig:RR_allz_sersic} (${\alpha \approx 0.029}$ in
Figure \ref{fig:RR_allz_ssfr}) and lie close to the $J/M$ equality line 
(within $\lesssim$ 0.1--0.2 dex) with no discernible evolution.
(There is a slight offset to smaller sizes in the late-type sample
when selected by sSFR rather than S\'ersic index.) This result agrees with recent direct
measurements of specific angular momentum at $0.2 < z < 1.4$ \citep{Contini:2016fn} and at
$1 < z < 3$ \citep{Burkert:2016fr}, which show that $J/M$ in galactic disks is
nearly the same as in their dark matter halos.

\cite{Kravtsov:2013cy} speculated that the sizes of galaxies grew in
proportion to the sizes of their halos until $z \sim 2$ and then stopped,
while their halos continued to grow in mass and size. 
We find instead that the $\Reff$--$\Rhalo$ relations at $z < 2$ are
very similar to those at $z > 2$. Our $\Reff$--$\Rhalo$ relations for the
late-type galaxies at $z < 0.5$ have smaller amplitudes than those at
$z > 0.5$, indicating a possible slowdown in the growth of disks, but this
deviation is mild ($\sim$0.2 dex) and not established beyond all doubt (see below).

The $\Reff$--$\Rhalo$ relation for early-type galaxies
is also nearly constant. We see 
in Figures \ref{fig:RR_allz_sersic} and \ref{fig:RR_allz_ssfr} that the 
trend for early-type galaxies at all redshifts roughly parallels that
for late-type galaxies, but shifted down by $\sim 0.2$ dex at $0 < z < 0.5$
and by $\sim$ 0.2--0.3 dex at $0.5 < z < 3$. There is a slight hint of a ``turnover''
at the most massive end at $0 < z < 0.5$ (see Figures \ref{fig:RR_allz_sersic} and
\ref{fig:RR_allz_ssfr}). This turnover, if real, could be due to either
size-measurement biases (due to diffuse
outer halos surrounding central galaxies in groups and clusters) or
the breakdown of abundance matching for the group- or cluster-mass halos.

\section{Uncertainties}\label{sec:errors}

How robust are these results?  The uncertainties in this study 
potentially include measurement and statistical errors internal to the 
CANDELS data set, as well as external systematic errors from the adopted
SMHM relations and stellar population models.  Here we provide a brief
assessment of these uncertainties.
 
As noted in Section \ref{sec:data}, errors in the measurements of 
effective radii $\Reff$ (from fits to S\'ersic profiles) are relatively 
small: $< 20\%$ (systematic) to $30\%$ (random). Even if these errors 
were at the upper end of this range for all galaxies and varied 
systematically with galactic masses and sizes, they would have a negligible 
influence on the coefficient and exponent of the galaxy size--halo size relation:
$\Reff = \alpha \Rhalo^{\beta}$ with $|\Delta\alpha/\alpha| \la 0.02$ and 
$|\Delta\beta| \la 0.08$ (assuming a $\sim 20\%$ or smaller systematic deviation 
in $\Reff$ for a factor of 10 or more variation in $\Rhalo$). Because the 
sample size in this study is so large ($N \sim 38000$), the effects of 
random errors in the size measurements on the mean $\Reff$--$\Rhalo$ relations 
are even smaller.  In a situation like this, with negligible internal errors, 
formal tests of goodness of fit are not informative, and we do not attempt them.

The dominant uncertainties in our galaxy size--halo size relations are most 
likely caused by possible systematic errors in our adopted SMHM relations. 
We can judge the magnitude of these errors by comparing the $\Reff$--$\Rhalo$ 
relations plotted in Figures \ref{fig:RR_z0} to \ref{fig:RR_z0_ssfr} for the
four different SMHM relations.  This comparison indicates that the SMHM relation 
may be responsible for systematic errors at the level of $\sim 0.1$--$0.2$ dex, 
perhaps a little less for the combined sample of galaxies, perhaps a little more 
for the subsamples split by morphological type.  Quantitative measures of the 
deviations among the $\Reff$--$\Rhalo$ relations at $0 < z < 0.5$ confirm these impressions.  

The contributions to the error budget from the adopted stellar population models, 
which determine the stellar masses and specific star formation rates, are 
smaller than those from the adopted SMHM relations.  Systematic errors in stellar 
masses could affect the $\Reff$--$\Rhalo$ relations at about the same level as 
systematic errors in $\Reff$.  The classification of the 3D shapes of galaxies 
(i.e., flat disks vs. round spheroids) by S\'ersic index is another source of uncertainty, 
because it is based only on the radial decline of the projected 2D surface brightness 
profiles. Fitting a single S\'ersic profile instead of a detailed disk/bulge
decomposition possibly adds further uncertainty.
Nevertheless, the $\Reff$--$\Rhalo$ relations we obtain from subsamples 
split by S\'ersic index agree at the $\lesssim 0.1$ dex level with those from 
subsamples split by specific star formation rate.

We estimate the impact of selection biases on our galaxy size--halo size
relations from the detection efficiencies for the CANDELS survey derived by
\cite{Guo:2013ig} as follows. They divide the $\Reff$--$H_{160}$ plane into regions
that are 0--50\%, 50--90\%, and 90--100\% complete. Most of our sample (88\%)
lies in the region of 90--100\% completeness, while the remainder (12\%) lies in the
region of 50--90\% completeness. To place an upper limit on the impact of
selection biases, we adopt the lower limits of 90\% and 50\% on the completeness
in these two regions of the $\Reff$--$H_{160}$ plane, assign weights 2.0 (i.e., $1/0.5$)
and 1.1 (i.e., $1/0.9$) to the galaxies in our sample in these regions, and then
recompute the $\Reff$--$\Rhalo$ relations. For $\Rhalo \gtrsim 100$ kpc, we
find negligible corrections to the median $\Reff$--$\Rhalo$ relations, while
for $\Rhalo \lesssim 100$ kpc, we find corrections below 0.1 dex for all galaxy types
and redshifts $0 < z < 3$. We conclude from this exercise that selection biases
are likely to be subdominant sources of uncertainty in our $\Reff$--$\Rhalo$
relations.

Based on this assessment of uncertainties, most of the results of this paper 
appear to be robust.  In particular, there is a strong, approximately linear 
correlation between the sizes of galaxies and their dark matter halos over the full 
range of redshifts examined here, $0 < z < 3$. The coefficient of proportionality 
is larger for late-type galaxies than for early-type galaxies, which follow roughly 
parallel sequences, except possibly at the highest redshifts. For late-type galaxies, 
the observed $\Reff$--$\Rhalo$ relation is generally consistent with simple models 
in which galactic disks grow with the same specific angular momentum as their dark matter halos.
There is some evidence for a slowdown in disk growth at $z < 0.5$, but the 
apparent deviation from the $J/M$ equality line is only $\sim 0.2$ dex.

\newpage
\floattable
\begin{deluxetable*}{lcccc}
\tabletypesize{\small}
\tablewidth{\textwidth}
\tablecaption{Verification of Main Results \label{tab:truth}}
\tablehead{\colhead{} & \colhead{SMHM 1} & \colhead{SMHM 2} & \colhead{SMHM 3} & \colhead{SMHM 4}}

\startdata
1. The $\Reff$--$\Rhalo$ relations are roughly linear in all redshift bins.  &  T  &  T  &  T  &  T \\
2. The $\Reff$--$\Rhalo$ relations are offset for early- and late-type galaxies.  &  T  &  T  &  T  &  T \\
3. The $\Reff$--$\Rhalo$ relation for late-type galaxies are close to the $J/M$ equality line. &  T  &  T  &  T  &  T \\
4. The $\Reff$--$\Rhalo$ relation shows little evolution between $z=0$ and $z=3$. &  T  &  T  &  T  &  T \\
\enddata

\end{deluxetable*}

We have plotted and examined the $\Reff$--$\Rhalo$ relations at all redshifts ($0 < z < 3$)
for all four SMHM relations to determine whether or not they support the four main results
discussed in Section \ref{sec:results}. The outcome of this test is recorded in Table \ref{tab:truth} by a T (for true) or F (for false) for each combination of 
SMHM relation and result. All of the entries are Ts. Table \ref{tab:truth}
therefore reinforces our conclusion that
the main scientific results of this study are robust relative to discrepancies among
the SMHM relations (because of the weak dependence of $\Rhalo$ on $\Mhalo$).


\section{Discussion}\label{sec:disc}

We have found that the sizes of galaxies are proportional on average to the
sizes of their dark matter halos over a wide range of galaxy and halo masses and over
the entire redshift range $0 < z < 3$ studied here: $\Reff=\alpha\Rhalo$ 
with $\alpha \approx 0.03$. In particular,
we confirm the basic relation found by \cite{Kravtsov:2013cy} at $z=0$ with only
minor adjustment, some of which is related to the difference between 2D 
half-light radii and 3D half-mass radii. There
is some curvature at the upper end of our overall $\Reff$--$\Rhalo$ relation, which
is due to the larger abundance and smaller average size of
early-type galaxies compared with late-type galaxies of the same stellar mass.  Indeed, we
find that early- and late-type galaxies follow distinct, roughly parallel 
$\Reff$--$\Rhalo$ relations offset by a factor of $\sim 2$ for the upper and lower
20th percentiles of S\'ersic index and specific star formation rate, 
which are meant to be proxies for disk-dominated and spheroid-dominated
galaxies.

Given the proportionality between galaxy and halo sizes, it is now straightforward to predict 
how galaxy sizes evolve with redshift, from the following alternative forms of equation (\ref{eq:rvir}): 
\begin{equation}\label{eq:size_evol}
\Reff=\alpha\Rhalo=\alpha\left[\frac{G\Mhalo}{100H^2(z)}\right]^{1/3}=\alpha\frac{{V_{200c}}}{10H(z)}.
\end{equation}
Here $H(z)$ is the Hubble parameter at redshift $z$, and ${V_{200c}}$ is the
circular velocity of the halo in question \citep[see][]{Mo:1998hg}. Thus, we
expect $\Reff\propto{H^{-2/3}(z)}$ or $\Reff\propto{H^{-1}(z)}$ depending on
whether galaxies at different $z$ are compared at the same $\Mhalo$ or ${V_{200c}}$.
As a result of gravitational clustering, the characteristic halo
mass evolves with redshift roughly as 
$\sigma({M^*_{200c}}, z) \propto \delta_c(z) / D(z)$, where $\sigma({M^*_{200c}}, z)$
is the RMS deviation of the linear density field smoothed over the scale ${R(M^*_{200c})}$,
$\delta_c(z)$ is the critical linear overdensity for collapse \citep{Kitayama:1996in},
and $D(z)$ is the linear growth factor \citep{Carroll:1992kw}.
The corresponding galactic size $\Reff^*(z)$ at the 
knee of the galaxy mass function should evolve 
according to equation (\ref{eq:size_evol}) with $\Mhalo \rightarrow {M^*_{200c}}(z)$.  
This expression for $\Reff^*(z)$ relates the typical
sizes of progenitor--descendant pairs of galaxies at different redshifts,
although there will be a large dispersion about it as a result of stochasticity
in the hierarchical growth of galaxies.

Our $\Reff$--$\Rhalo$ relations for late-type galaxies (defined by low $n$, high
sSFR) at $0.5 < z < 3$ are within $\lesssim$ 0.1--0.2 dex of the predictions of simple
models in which galactic disks acquire and retain the same specific angular
momentum as induced by tidal torques in their surrounding dark matter halos. At
$z < 0.5$, late-type galaxies are $\sim 0.2$ dex below this prediction. However,
given possible systematic errors in the measurements of galactic sizes
($\lesssim20\%$ for low-$n$ galaxies), our results are consistent with a range
$\eta_j \sim 80\%\pm20\%$ for the retained fraction of specific angular momentum. 
Our results therefore agree nicely with recent, direct
measurements of the specific angular momentum of galactic disks at $z=0$
\citep{Fall:2013du}, at $0.2<z<1.4$ \citep{Contini:2016fn}, and at $1 < z < 3$
\citep{Burkert:2016fr}, all of which indicate retention factors $\eta_j$ near unity or
slightly below.

The notion of angular momentum conservation was introduced as a simplifying
approximation in the era of analytical models of galaxy formation
\citep{Fall:1980up}.  Since then, hydrodynamical models have revealed a much
more complex situation.  In particular, it is now clear that several physical
processes may change the specific angular momentum of galaxies or
parts of galaxies during their formation and evolution, including merging,
feedback, inflows, outflows, and gravitational interactions between baryons and
dark matter.  Some of these processes cause gains in specific angular momentum,
while others cause losses (see \citealt{Romanowsky:2012kb} and
\citealt{Genel:2015kp} for summaries and references to earlier work).

The galactic disks that form in recent hydrodynamical simulations have nearly
the same specific angular momentum on average as their dark matter halos, in good
agreement with observations
\citep{Genel:2015kp,Pedrosa:2015dh,Teklu:2015ev,Zavala:2016ki}.  Evidently, the
processes responsible for gains and losses are either weak or in rough balance,
leading to an apparent (if not strict) conservation of angular momentum during
the formation of galactic disks. Simulations and now observations indicate that
galaxies of all types grow in a quasi-homologous (or self-similar) relationship
with their dark matter halos. The details of how this happens are a topic of ongoing research.\\

We thank Gerard Lemson for the help with Millennium Simulation, Adam Tomczak
for useful discussions of stellar mass functions, and Andrey Kravtsov for
providing conversion factors between different halo mass definitions.
We also thank Avishai Dekel, Sandra Faber, Steve Finkelstein, 
Andrey Kravtsov, Yu Lu, and Rachel Somerville for comments 
on a near-final draft of this paper. This work is based on observations taken by the 
CANDELS Multi-Cycle Treasury Program with the NASA/ESA HST, 
which is operated by the Association of Universities for Research in Astronomy, Inc.,
under NASA contract NAS5-26555.

\appendix

\section*{Transformation Between the $\Reff$--$M_*$ and $\Reff$--$\Rhalo$ Relations}\label{sec:appendix}\label{sec:appendix}

The halo virial radius $\Rhalo$ of each galaxy in our sample was computed by 
the abundance-matching technique, i.e., from its stellar mass $M_*$, the SMHM 
relation, and Equation (\ref{eq:rvir}).  
Thus, the positions of galaxies in the $\Reff$--$\Rhalo$ plane represent a 
nonlinear transformation of their positions in the $\Reff$--$M_*$ plane.  
While the former is more fundamental from a theoretical perspective and is 
the main focus of this paper, the latter is one step closer to the observations, 
since it requires only the conversion of luminosities and colors into stellar 
masses.  
It is therefore of interest to examine the $\Reff$--$M_*$ diagrams for our sample 
and how they map into the $\Reff$--$\Rhalo$ diagrams presented in Section \ref{sec:results}.
This is the purpose of this appendix.

\begin{figure*}[ht]
\figurenum{10}
\plotone{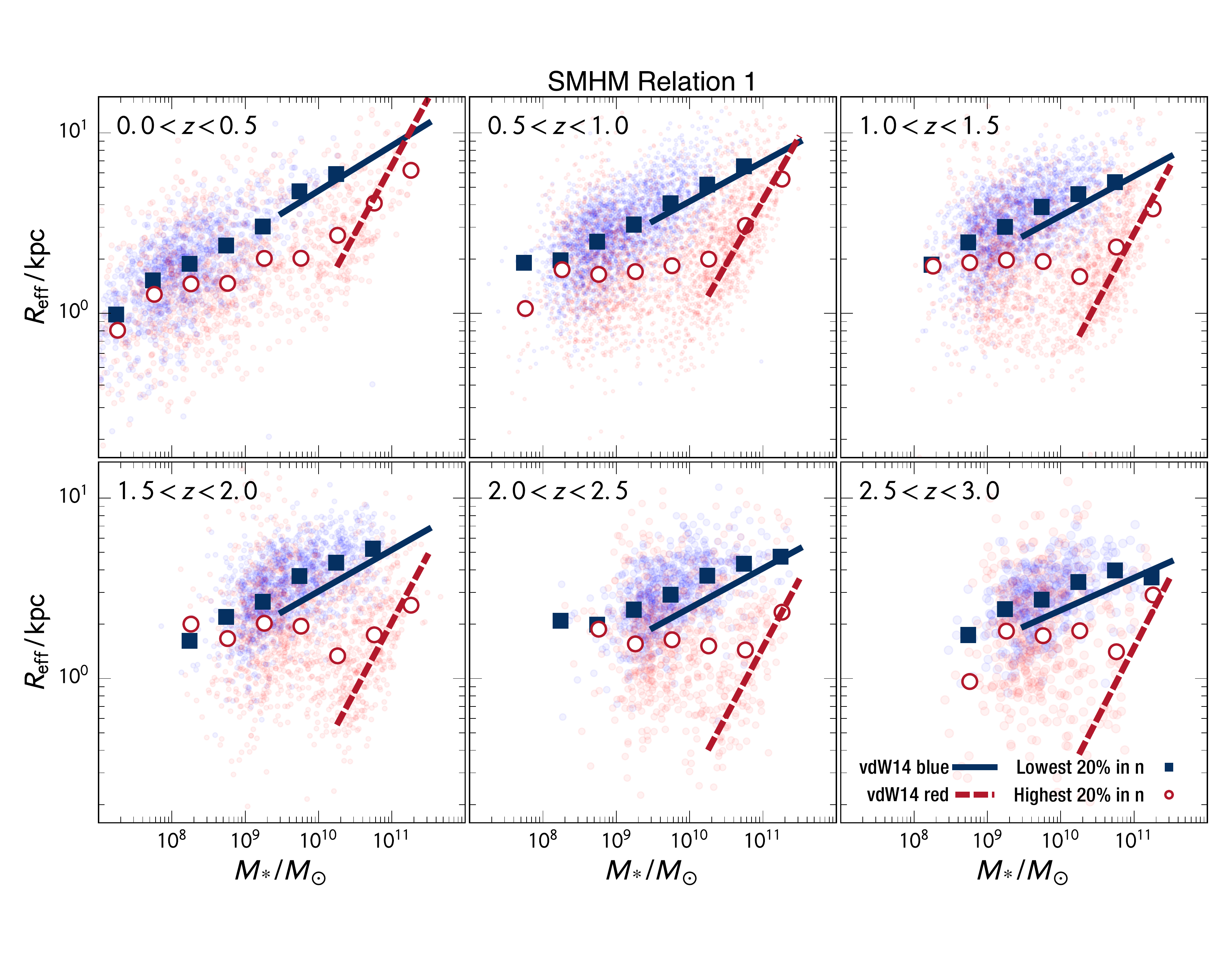}
\figcaption{Galaxy effective radius $\Reff$ plotted against stellar mass $M_*$ at
different redshifts for subsamples of galaxies with the lowest and highest 20\% 
of the measured S\'ersic index $n$ as proxies for late- and early-type 
galaxies, respectively.
The six panels show results computed from SMHM relation 1
in redshift intervals of $\Delta z = 0.5$ covering the range $0 < z < 3$.
The faint blue and red dots represent individual low-$n$ and high-$n$ galaxies, 
respectively, while the filled blue squares and open red circles indicate the 
corresponding median values of $\Reff$ in bins of width 0.5 in $\log M_*$.
The blue solid and red dashed lines show the power-law fits to the  
$\Reff$--$M_*$ relations for blue and red galaxies (defined in terms of
rest-frame $UVJ$ colors) from \cite{vanderWel:2014hi}.
Note that our sample extends to fainter and therefore less massive galaxies 
than the \cite{vanderWel:2014hi} sample. Compare with Figure \ref{fig:RR_comp}.
\label{fig:RM_comp}}
\end{figure*}

Figure \ref{fig:RM_comp} shows the $\Reff$--$M_*$ diagram for galaxies in our sample in 
six redshift intervals covering the range $0<z<3$ when divided, as before, 
into subsamples with the lowest and highest quintiles of S\'ersic index $n$.  
We also plot in this diagram 
the median values of $\Reff$ in bins of width 0.5 in $\log M_*$ for these two 
subsamples.  Evidently, the median $\Reff$--$M_*$ relation for low-$n$ 
galaxies is close to a single power law (a straight line in a plot of $\log\Reff$ 
against $\log M_*$), whereas the relation for high-$n$ galaxies is more complicated: 
it is flatter than the low-$n$ relation at low masses and steeper at high masses, 
with a bend at $M_* \sim{\rm few} \times10^{10}~M_\odot$. 
It is also clear from Figure \ref{fig:RM_comp} that the median $\Reff$--$M_*$ relations for 
both low-$n$ and high-$n$ galaxies evolve very slowly.
For subsamples with the highest and lowest quintiles of specific star formation rate,
we find similar behaviors in the median $\Reff$--$M_*$ relations, as functions 
of both $M_*$ and $z$, especially for $z < 1.5$ (not shown here).

\begin{figure*}[h]
\figurenum{11}
\plotone{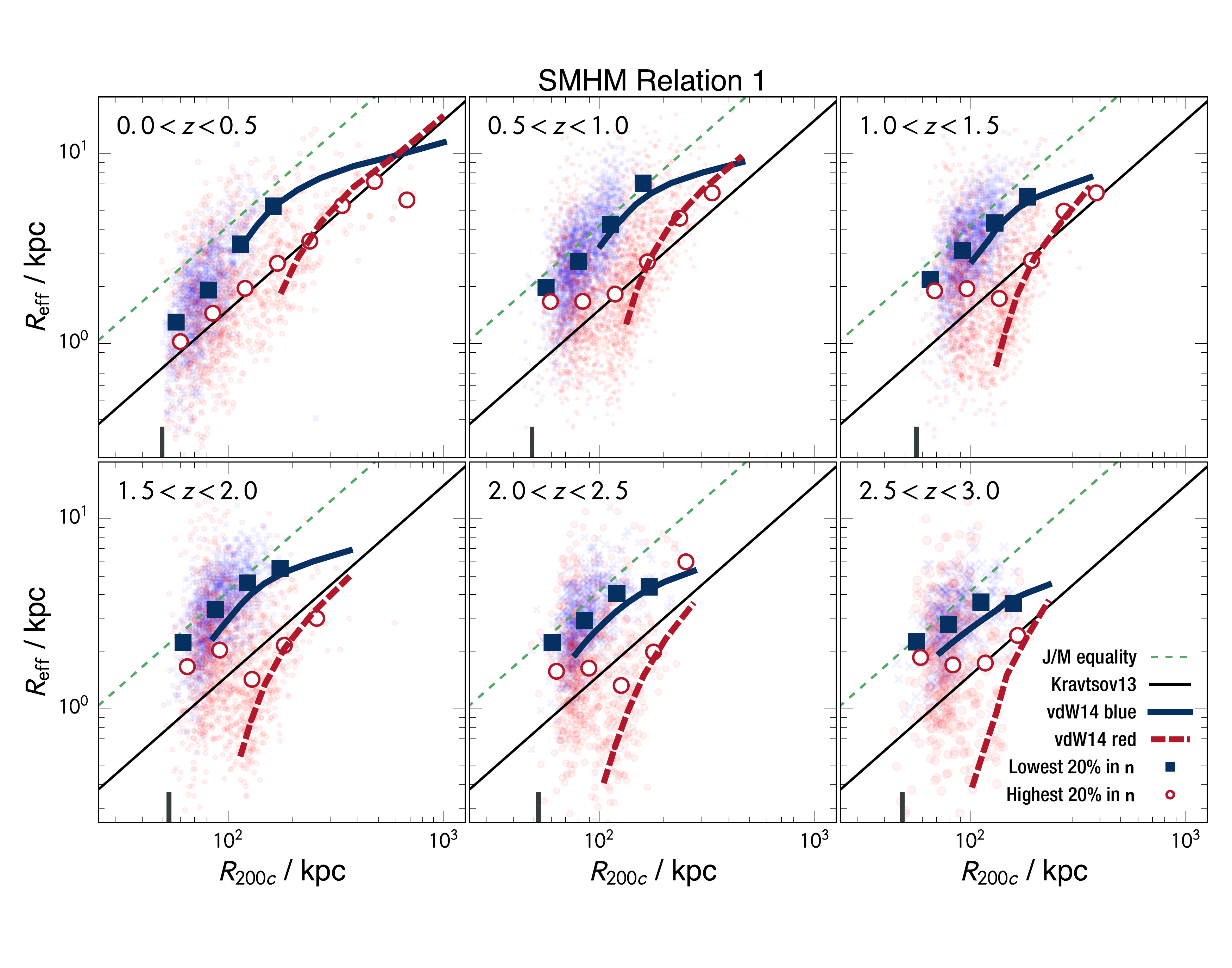}
\figcaption{Galaxy effective radius $\Reff$ plotted against halo virial radius $\Rhalo$ at
different redshifts for subsamples of galaxies with the lowest and highest 20\% 
of the measured S\'ersic index $n$ as proxies for late- and early-type 
galaxies, respectively.
The six panels show results computed from SMHM relation 1
in redshift intervals of $\Delta z = 0.5$ covering the range $0 < z < 3$.
The faint blue and red dots represent individual low-$n$ and high-$n$ galaxies, 
respectively, while the filled blue squares and open red circles indicate the 
corresponding median values of $\Reff$ in bins of width 0.15 in $\log\Rhalo$.
The diagonal solid lines show the $R_{1/2}$--$\Rhalo$ relation at $z=0$ from 
\cite{Kravtsov:2013cy} assuming $\Reff=R_{1/2}$, while the diagonal dashed 
lines show the prediction for galactic disks with the same $J/M$ as
their surrounding halos.
The thick tick mark at the bottom of each panel indicates the halo mass 
corresponding to the reference stellar mass $M_{*,\rm{low}}$ listed in 
Table \ref{tab:sample}.
The blue solid and red dashed curves are the power-law fits for blue 
and red galaxies in the $\Reff$--$M_*$ plane after transformation into 
the $\Reff$--$\Rhalo$ plane. Compare with Figure \ref{fig:RM_comp}.
\label{fig:RR_comp}}
\end{figure*}

Figure \ref{fig:RR_comp} shows the result of transforming the $\Reff$--$M_*$ diagram into
the $\Reff$--$\Rhalo$ diagram with SMHM relation 1.  This is exactly the 
same as Figure \ref{fig:RR_allz_sersic} except that we have omitted the vertical bars for clarity.  
We have already discussed this diagram at length in Section \ref{sec:results}.
Here we note only that the median $\Reff$--$\Rhalo$ relations for low-$n$ 
and high-$n$ galaxies in Figure \ref{fig:RR_comp} appear more parallel than the corresponding 
$\Reff$--$M_*$ relations in Figure \ref{fig:RM_comp}, particularly at $z < 1.5$, where they are
best defined.  
This is a consequence of the nonlinearity of the SMHM relation, 
especially near $\Mhalo \sim10^{12}~M_\odot$, corresponding to 
$M_* \sim{\rm few} \times10^{10}~M_\odot$, and hence near the bend in 
the $\Reff$--$M_*$ relation for high-$n$ galaxies.

\cite{vanderWel:2014hi} also derived $\Reff$--$M_*$ relations 
in the redshift range $0<z<3$ for galaxies in the CANDELS sample.
The main difference between their work and ours is that they 
adopted the same selection limits in all CANDELS regions, whereas we
adopted fainter selection limits in the Deep and HUDF regions.
As a result, our $\Reff$--$M_*$ relations extend to much lower
$M_*$ than theirs.  
Otherwise, the selection of galaxies and measurement of their properties 
are nearly identical in the two studies.
\cite{vanderWel:2014hi} divided their sample into blue and red galaxies
on the basis of rest-frame $UVJ$ colors rather than by S\'ersic index or
specific star formation rate, as we have done.  
Naturally, there is a general, but not a perfect, correspondence 
between these three different proxies for late- and early-type galaxies.

\cite{vanderWel:2014hi} fitted power laws to the $\Reff$--$M_*$ relations
for blue and red galaxies; these are shown in Figure \ref{fig:RM_comp} as the blue solid 
and red dashed line segments, respectively.
For red galaxies, they truncated the fits at $M_* = 2 \times10^{10}~M_\odot$ 
because they also noticed a bend in the $\Reff$--$M_*$ relation near this mass
and a flattening below it.
We obtain nearly identical results when we divide our sample into blue and red 
galaxies using the same cuts in rest-frame $UVJ$ colors as \cite{vanderWel:2014hi}.
The blue solid and red dashed curves in Figure \ref{fig:RR_comp} show how the 
\cite{vanderWel:2014hi} power laws in the $\Reff$--$M_*$ diagram transform into the 
$\Reff$--$\Rhalo$ diagram.
As expected, this mapping introduces curvature and makes the $\Reff$--$\Rhalo$
relations for blue and red galaxies somewhat more parallel. 
However, the transformed relations cover only a narrow range of halo sizes,
roughly $100~{\rm kpc} \lesssim \Rhalo \lesssim 300~{\rm kpc}$, except in the lowest
redshift interval.
We have been able to extend the $\Reff$--$\Rhalo$ relations to a wider range 
of halo sizes, roughly ${50}~{\rm kpc} \lesssim \Rhalo \lesssim 300~{\rm kpc}$,
with our fainter selection limits in the CANDELS Deep and HUDF regions.

\begin{thebibliography}{}

\bibitem[Behroozi et al.(2013)]{Behroozi:2013fg} Behroozi, P.~S., Wechsler, R.~H., \& Conroy, C.\ 2013, \apj, 770, 57 
\bibitem[Bertin \& Arnouts(1996)]{Bertin:1996ww} Bertin, E., \& Arnouts, S.\ 1996, \aaps, 117, 393 
\bibitem[Bett et al.(2007)]{2007MNRAS.376..215B} Bett, P., Eke, V., Frenk, C.~S., et al.\ 2007, \mnras, 376, 215 
\bibitem[Boylan-Kolchin et al.(2009)]{BoylanKolchin:2009co} Boylan-Kolchin, M., Springel, V., White, S.~D.~M., Jenkins, A., \& Lemson, G.\ 2009, \mnras, 398, 1150
\bibitem[Bouwens et al.(2001)]{Bouwens:2010a} Bouwens, R., Illingworth, G.~D., Oesch, P.~A., et al.\ 2010, \apj, 709, L133
\bibitem[Bouwens et al.(2010)]{Bouwens:2010dk} Bouwens, R.~J., Illingworth, G.~D., Oesch, P.~A., et al.\ 2010, \apjl, 709, L133
\bibitem[Bryan \& Norman(1998)]{Bryan:1998cc} Bryan, G.~L., \& Norman, M.~L.\ 1998, \apj, 495, 80 
\bibitem[Bullock et al.(2001)]{Bullock:2001kb} Bullock, J.~S., Dekel, A., Kolatt, T.~S., et al.\ 2001, \apj, 555, 240 
\bibitem[Burkert et al.(2016)]{Burkert:2016fr} Burkert, A., F{\"o}rster Schreiber, N.~M., Genzel, R., et al.\ 2016, \apj, 826, 214 

\bibitem[Carroll et al.(1992)]{Carroll:1992kw} Carroll, S.~M., Press, W.~H., \& Turner, E.~L.\ 1992, \araa, 30, 499 
\bibitem[Chabrier(2003)]{Chabrier:2003ki} Chabrier, G.\ 2003, \pasp, 115, 763 
\bibitem[Cole et al.(2000)]{Cole:2000fl} Cole, S., Lacey, C.~G., Baugh, C.~M., \& Frenk, C.~S.\ 2000, \mnras, 319, 168 
\bibitem[Contini et al.(2016)]{Contini:2016fn} Contini, T., Epinat, B., Bouch{\'e}, N., et al.\ 2016, \aap, 591, A49
\bibitem[Croton et al.(2016)]{Croton:2016jg} Croton, D.~J., Stevens, A.~R.~H., Tonini, C., et al.\ 2016, \apjs, 222, 22  
\bibitem[Curtis-Lake et al.(2016)]{CurtisLake:2016bm} Curtis-Lake, E., McLure, R.~J., Dunlop, J.~S., et al.\ 2016, \mnras, 457, 440 

\bibitem[Diemer et al.(2013)]{Diemer:2013hu} Diemer, B., More, S., \& Kravtsov, A.~V.\ 2013, \apj, 766, 25 
\bibitem[Diemer \& Kravtsov(2015)]{Diemer:2015bd} Diemer, B., \& Kravtsov, A.~V.\ 2015, \apj, 799, 108 
\bibitem[Dutton et al.(2010)]{Dutton:2010hf} Dutton, A.~A., Conroy, C., van den Bosch, F.~C., Prada, F., \& More, S.\ 2010, \mnras, 407, 2 

\bibitem[Ellis et al.(2013)]{2013ApJ...763L...7E} Ellis, R.~S., McLure, R.~J., Dunlop, J.~S., et al.\ 2013, \apjl, 763, L7 

\bibitem[Fall \& Efstathiou(1980)]{Fall:1980up} Fall, S.~M., \& Efstathiou, G.\ 1980, \mnras, 193, 189 
\bibitem[Fall(1983)]{Fall:1983wu} Fall, S.~M.\ 1983, in IAU Symp. 100, Internal Kinematics and Dynamics of Galaxies, ed. E. Athanassoula (Cambridge: Cambridge Univ. Press), 391 
\bibitem[Fall \& Romanowsky(2013)]{Fall:2013du} Fall, S.~M., \& Romanowsky, A.~J.\ 2013, \apjl, 769, L26 
\bibitem[Ferguson et al.(2004)]{Ferguson:2004dt} Ferguson, H.~C., Dickinson, M., Giavalisco, M., et al.\ 2004, \apjl, 600, L107 

\bibitem[Galametz et al.(2013)]{Galametz:2013dd} Galametz, A., Grazian, A., Fontana, A., et al.\ 2013, \apjs, 206, 10 
\bibitem[Genel et al.(2015)]{Genel:2015kp} Genel, S., Fall, S.~M., Hernquist, L., et al.\ 2015, \apjl, 804, L40 
\bibitem[Grogin et al.(2011)]{Grogin:2011hx} Grogin, N.~A., Kocevski, D.~D., Faber, S.~M., et al.\ 2011, \apjs, 197, 35 
\bibitem[Guo et al.(2013)]{Guo:2013ig} Guo, Y., Ferguson, H.~C., Giavalisco, M., et al.\ 2013, \apjs, 207, 24 


\bibitem[Hathi et al.(2008)]{Hathi:2008ca} Hathi, N.~P., Malhotra, S., \& Rhoads, J.~E.\ 2008, \apj, 673, 686-693 
\bibitem[Huang et al.(2013)]{Huang:2013kb} Huang, K.-H., Ferguson, H.~C., Ravindranath, S., \& Su, J.\ 2013, \apj, 765, 68
\bibitem[Hudson et al.(2015)]{2015MNRAS.447..298H} Hudson, M.~J., Gillis, B.~R., Coupon, J., et al.\ 2015, \mnras, 447, 298 

\bibitem[Illingworth et al.(2013)]{2013ApJS..209....6I} Illingworth, G.~D., Magee, D., Oesch, P.~A., et al.\ 2013, \apjs, 209, 6 

\bibitem[Kitayama \& Suto(1996)]{Kitayama:1996in} Kitayama, T., \& Suto, Y.\ 1996, \apj, 469, 480 
\bibitem[Koekemoer et al.(2011)]{Koekemoer:2011br} Koekemoer, A.~M., Faber, S.~M., Ferguson, H.~C., et al.\ 2011, \apjs, 197, 36 
\bibitem[Koekemoer et al.(2013)]{2013ApJS..209....3K} Koekemoer, A.~M., Ellis, R.~S., McLure, R.~J., et al.\ 2013, \apjs, 209, 3 
\bibitem[Kravtsov(2013)]{Kravtsov:2013cy} Kravtsov, A.~V.\ 2013, \apjl, 764, L31 

\bibitem[Mandelbaum et al.(2016)]{Mandelbaum:2016eb} Mandelbaum, R., Wang, W., Zu, Y., et al.\ 2016, \mnras, 457, 3200 
\bibitem[Mo et al.(1998)]{Mo:1998hg} Mo, H.~J., Mao, S., \& White, S.~D.~M.\ 1998, \mnras, 295, 319 
\bibitem[Mobasher et al.(2015)]{Mobasher:2015gp} Mobasher, B., Dahlen, T., Ferguson, H.~C., et al.\ 2015, \apj, 808, 101 
\bibitem[Moustakas et al.(2013)]{Moustakas:2013il} Moustakas, J., Coil, A.~L., Aird, J., et al.\ 2013, \apj, 767, 50 
\bibitem[Mosleh et al.(2012)]{Mosleh:2012cw} Mosleh, M., Williams, R.~J., Franx, M., et al.\ 2012, \apjl, 756, L12 

\bibitem[Nayyeri et al.(2016)]{2016arXiv161207364N} Nayyeri, H., Hemmati, S., Mobasher, B., et al.\ 2016, arXiv:1612.07364 

\bibitem[Pedrosa \& Tissera(2015)]{Pedrosa:2015dh} Pedrosa, S.~E., \& Tissera, P.~B.\ 2015, \aap, 584, A43 
\bibitem[Peng et al.(2010)]{Peng:2010eh} Peng, C.~Y., Ho, L.~C., Impey, C.~D., \& Rix, H.-W.\ 2010, \aj, 139, 2097 

\bibitem[Rodr{\'{\i}}guez-Puebla et al.(2015)]{RodriguezPuebla:2015bk} Rodr{\'{\i}}guez-Puebla, A., Avila-Reese, V., Yang, X., et al.\ 2015, \apj, 799, 130 
\bibitem[Romanowsky \& Fall(2012)]{Romanowsky:2012kb} Romanowsky, A.~J., \& Fall, S.~M.\ 2012, \apjs, 203, 17 

\bibitem[Salmon et al.(2015)]{Salmon:2015iz} Salmon, B., Papovich, C., Finkelstein, S.~L., et al.\ 2015, \apj, 799, 183 
\bibitem[Santini et al.(2015)]{Santini:2015hh} Santini, P., Ferguson, H.~C., Fontana, A., et al.\ 2015, \apj, 801, 97 
\bibitem[Shibuya et al.(2015)]{Shibuya:2015bj} Shibuya, T., Ouchi, M., \& Harikane, Y.\ 2015, \apjs, 219, 15 
\bibitem[Straatman et al.(2016)]{2016ApJ...830...51S} Straatman, C.~M.~S., Spitler, L.~R., Quadri, R.~F., et al.\ 2016, \apj, 830, 51 
\bibitem[Szomoru et al.(2013)]{Szomoru:2013gh} Szomoru, D., Franx, M., van Dokkum, P.~G., et al.\ 2013, \apj, 763, 73 

\bibitem[Taghizadeh-Popp et al.(2015)]{TaghizadehPopp:2015hp} Taghizadeh-Popp, M., Fall, S.~M., White, R.~L., \& Szalay, A.~S.\ 2015, \apj, 801, 14 
\bibitem[Teklu et al.(2015)]{Teklu:2015ev} Teklu, A.~F., Remus, R.-S., Dolag, K., et al.\ 2015, \apj, 812, 29 
\bibitem[Tomczak et al.(2014)]{Tomczak:2014hw} Tomczak, A.~R., Quadri, R.~F., Tran, K.-V.~H., et al.\ 2014, \apj, 783, 85 

\bibitem[van der Wel et al.(2012)]{vanderWel:2012eu} van der Wel, A., Bell, E.~F., H{\"a}ussler, B., et al.\ 2012, \apjs, 203, 24 
\bibitem[van der Wel et al.(2014)]{vanderWel:2014hi} van der Wel, A., Franx, M., van Dokkum, P.~G., et al.\ 2014, \apj, 788, 28 



\bibitem[Zavala et al.(2016)]{Zavala:2016ki} Zavala, J., Frenk, C.~S., Bower, R., et al.\ 2016, \mnras, 460, 4466
\end{thebibliography}

\end{document}